\documentclass[fleqn,usenatbib]{mnras}

\usepackage{newtxtext,newtxmath}
\usepackage[T1]{fontenc}
\usepackage{ae,aecompl}

\usepackage{graphicx}	% Including figure files
\usepackage{amsmath}	% Advanced maths commands
\usepackage{pdflscape}
\usepackage{afterpage}
\usepackage{rotating}

\newcommand{\chandra}{{\it Chandra}}
\newcommand{\nustar}{{\it NuSTAR}}

\newcommand{\swift}{{\it Swift}}
\newcommand{\xmm}{{\it XMM-Newton}}

\newcommand{\ergps}{erg~cm$^{-2}$~s$^{-1}$}
\newcommand{\blue}{\textcolor{black}}
\newcommand{\srcfull}{KUG~1141$+$371}
\newcommand{\src}{KUG~1141}
\title[Multi-wavelength variability of \srcfull]{The Awakening Beast in the Seyfert 1 Galaxy \srcfull\ I. }

\author[J. Jiang et al.]{Jiachen Jiang,$^{1,2}$\thanks{E-mail: jcjiang@tsinghua.edu.cn} 
Huaqing Cheng,$^{12}$
Luigi C. Gallo,$^{3}$ 
Luis C. Ho,$^{4,5}$ 
Douglas J. K. Buisson,$^{6}$ \newauthor 
Andrew C. Fabian,$^{7}$
Fiona A. Harrison,$^{8}$
Michael L. Parker,$^{9}$ 
Christopher S. Reynolds,$^{7}$ \newauthor
James F. Steiner,$^{10}$
John A. Tomsick$^{11}$,
Dominic J. Walton$^{7}$ and
Weimin Yuan$^{12}$
\\
\\
$^{1}$Department of Astronomy, Tsinghua University, Shuangqing Road 30, Beijing 100034, China\\
$^{2}$Tsinghua Center for Astrophysics, Tsinghua University, Shuangqing Road 30, Beijing 100034, China\\
$^{3}$Department of Astronomy and Physics, Saint Mary's University, 923 Robie Street, Halifax, NS, B3H 3C3, Canada \\
$^{4}$Kavli Institute for Astronomy and Astrophysics, Peking University, Beijing 100871, China\\
$^{5}$Department of Astronomy, School of Physics, Peking University, Beijing 100871, China\\
$^{6}$Department of Astronomy,University of Southampton, Southampton S017 1BJ, UK\\
$^{7}$Institute of Astronomy, University of Cambridge, Madingley Road, Cambridge CB3 0HA, UK\\
$^{8}$Cahill Center for Astronomy and Astrophysics, California Institute of Technology, Pasadena, CA 91125, USA\\
$^{9}$European Space Agency, European Space Astronomy Centre, E-28691 Villanueva de la Ca\~{n}ada, Madrid, Spain\\
$^{10}$Harvard-Smithsonian Center for Astrophysics, 60 Garden Street, Cambridge, MA 02138, USA\\
$^{11}$Space Sciences Laboratory, 7 Gauss Way, University of California, Berkeley, CA 94720-7450, USA\\
$^{12}$National Astronomical Observatory of China, 20A Datun Road, Chaoyang District, Beijing, China\\
}

\date{Accepted XXX. Received YYY; in original form ZZZ}

\pubyear{2020}

\begin{document}
\label{firstpage}
\pagerange{\pageref{firstpage}--\pageref{lastpage}}
\maketitle

\begin{abstract}
\srcfull\ is a Seyfert 1 galaxy that shows a simultaneous flux increase in the optical and UV bands in the past decade. For instance, the latest \swift\ observation in 2019 shows that the UVW2 flux of the AGN in \srcfull\ has increased by over one order of magnitude since 2009. Meanwhile, the soft X-ray flux of \srcfull\ also shows a steady increase by one order of magnitude since 2007. \blue{The significant multi-wavelength luminosity change is likely due to a boost of mass accretion rate from approximately $0.6\%$ of the Eddington limit to $3.2\%$,} assuming a black hole mass of $10^{8}$\,$M_{\odot}$. In this work, we conduct detailed multi-epoch X-ray spectral analysis focusing on the variability of the X-ray continuum emission and the puzzling soft excess emission. In addition, our SED models also suggest a simultaneous increase of disc temperature and a decreasing inner disc radius along with the increasing accretion rate. Finally, we discuss possible connection between \srcfull\ and black hole transients in outburst.   
\end{abstract}

\begin{keywords}
accretion, accretion discs\,-\,black hole physics, X-ray: galaxies, galaxies: Seyfert
\end{keywords}

%%%%%%%%%%%%%%%%%%%%%%%%%%%%%%%%%%%%%%%%%%%%%%%%%%

%%%%%%%%%%%%%%%%% BODY OF PAPER %%%%%%%%%%%%%%%%%%

\section{Introduction}

The nuclei in Seyfert galaxies \citep[Sys,][]{seyfert43} are very strong and variable emitters. The optical variability of most sources is consistent with small stochastic variability \citep[e.g.][]{kelly09, macleod10}. Such variability might be related to the reprocessing of rapidly changing X-ray emission from a more compact region near the central black hole \citep[BH, e.g.][]{clavel92,uttley13,buisson17,gallo18}. However, more significant magnitude change is often seen in some flaring or dimming Sys \citep[e.g.][]{khachikyan71,tohline76,penston84,cohen86}, which cannot be explained by reprocessing. Their optical emission is found to change significantly within a much shorter time interval than the viscous timescales given by the standard thin disc model, e.g. approximately $10^{5}$ years for R=100\,$r_{\rm g}$ around a $10^{8}$\,$M_{\odot}$ BH. Some recent studies suggest that the optical continuum emission may originate from even larger radii \citep[e.g.][]{shappee14,troyer16}, corresponding to even longer timescales. Some of these peculiar Sys also switch look between Sy1 and Sy2 by presenting simultaneous appearance or disappearance of broad optical permitted lines along with magnitude changes \citep[e.g.][]{runnoe16}.

Two scenarios have been proposed to explain these rapidly changing Sys: a variable line-of-sight column density due to the clumpiness of the torus and a sudden change in the mass accretion rate. In the former case, our line of sight towards the source may intercept with the edge of the torus. A rapid change in the observed magnitude can happen when a moving clump comes across the line of sight \citep[e.g.][]{goodrick89,leighly15}. The `hide-and-seek' of the broad emission lines can also be explained by such variable obscuration. In the latter case, a sudden change of the mass accretion rate in the disc may form or disrupt the broad line region. The broad permitted lines are photoionised by the continuum emission. Below a certain critical accretion rate/luminosity, no broad line regions can be formed, and thus the source shows disappearance of broad lines along with a dimming continuum emission, or vice verse \citep[e.g.][]{runnoe16, marin17}.  

Optical polarisation is one of the most efficient ways to disentangle the two scenarios mentioned above. The presence of broad emission lines in the polarised optical spectra of many Sy2s suggest the existence of a broad line region in Sy2s as in Sy1s \citep[e.g.][]{moran00}. \blue{If a high level of polarisation is still shown in a dimming Sy, the broad line region might still exist \citep[e.g.][]{marin17}.} Radiation from the AGN is scattered on the inner edge of the torus, and a large polarisation degree may be detected. The observed magnitude change is therefore due to variable obscuration along the line of sight. On the contrary, a rapid decrease of accretion rate may be responsible for the disappearance of the broad line region if the exhibited polarisation is low in the optical band \citep[e.g.][]{runnoe16,hutsemekers17}. The other approach of studying these unusual AGN is to search for evidence of simultaneous X-ray variability. X-rays are less affected by photoelectric absorption in neutral material than longer wavelengths. Therefore, X-ray spectra analysis will enable us to identify any variable line-of-sight obscuration \citep[e.g.][]{ricci16} or any changes in the intrinsic AGN spectrum due to a varying accretion rate \citep[e.g. soft excess,][]{noda18}.

In this paper, we focus on the long term multi-wavelength variability of an AGN called \srcfull\ (\src\ hereafter), which is poorly studied in the X-ray band. \src\ (z=0.038) is a Seyfert 1 galaxy \citep{runco16} with no evidence for nearby galaxy companions \citep{koss12}. \citet{oh15} studied the optical H$\alpha$ emission line of \src\ and estimated the mass of the central SMBH to be around $\log(M_{\rm BH}/M_{\odot})=7.99\pm0.06$ (statistical error) by following the method in \citet{greene05}. No obvious radio emission was found from \src\ in the Very Large Array (VLA) FIRST Survey \citep{wadadekar04}, indicating a radio-quiet nature of this source.

In Section\,\ref{reduction}, we introduce the data reduction process. In Section\,\ref{variability}, we give an overview of the long-term flux increase in \src\ in multiple wavelengths. In Section\,\ref{xray}, we conduct detailed X-ray spectral analysis. In Section\,\ref{sed}, we build SED models for \src. In Section\,\ref{discuss}, we discuss our results. In \blue{Appendix\,\ref{sec:uvoterror}, we estimate the uncertainty of the AGN flux of \src\ in the optical band.} In Appendix\,\ref{extra}, additional information of our X-ray spectral analysis is shown. In Appendix\,\ref{short}, short-term lightcurves of \src\ obtained by \xmm\ and \nustar\ are presented.

\section{Data Reduction} \label{reduction}

A list of X-ray observations of \src\ in the archive is shown in Table\,\ref{obs_log}. They are numbered by dates. There is only one \xmm\ observation (obs 5) and one \nustar\ observation (obs 12) in the archive for \src. A simultaneous \swift\ snapshot observation is analysed together with the \nustar\ observation.

\subsection{\xmm}

\subsubsection{EPIC}
We reduce \xmm\ data using the \xmm\ Science Analysis Software (SAS, v.18.0.0) and calibration files (ccf, v.20190513). The cleaned EPIC calibrated event lists are generated by using EMPROC and EPPROC for EPIC-MOS and EPIC-pn data respectively. We filter out the intervals that are dominated by flaring particle background. These high background intervals are defined as the periods when the single event count rate >10\,keV is higher than 0.35\,cts\,s$^{-1}$ for EPIC-MOS observations, and the single event count rate in the 10--12\,keV band is higher than 0.4\,cts\,s$^{-1}$ for the EPIC-pn observation. We then extract EPIC-MOS and EPIC-pn spectra by selecting single and double events from a circular region with radius of 20\,arcsec for EPIC-MOS data and 25\,arcsec for EPIC-pn. No evidence of pile-up effects is found by running the EPATPLOT tool. A polygon background region on the same chip is used to extract background spectrum for EPIC-pn, avoiding the areas dominated by background Cu~K emission from the underlying electronic circuits. A circular region of 50\,arcsec is used to extract background spectra for EPIC-MOS. The redistribution matrix and ancillary response files are generated by running the RMFGEN and ARFGEN tools. The EPIC spectra are grouped to have a minimum signal-to-noise of 6 and oversample the spectral resolution by a factor of 3, and are modelled over 0.5--10\,keV. 

\subsubsection{OM}
The Optical Monitor (OM) data are extracted using the task OMICHAIN. \blue{As part of the OMICHAIN task, the source and background regions given by the OMDETECT function are used. The source region is a circle around the source with radius of 5 arcsec. The background is subtracted by OMDETECT using the pixels within an annulus region. The inner and outer diameters of the annulus are 37 and 42 pixels\footnote{The size of each pixel is 0.48 arcsecond.}.} Only UVW1 and UVM2 filters were used during the \xmm\ observation of \src.  We use the OM2PHA tool to convert the photometry data to the OGIP Type II data format in order to be used in XSPEC. The response files for OM are downloaded from the SAS website\footnote{ftp://xmm.esac.esa.int/pub/ccf/constituents/extras/responses/OM}.

\subsection{\nustar}
We use the standard pipeline NUPIPELINE v.0.4.6 in HEASOFT v.6.26.1 to reduce the \nustar\ data. The calibration file version is v.20181030. A circular region with radii of 50\,arcsec is used to extract source spectra. The background spectra are extracted from the remaining regions on the same chip. The spectra are generated using the NUPRODUCTS tool. The FPMA and FPMB spectra are both grouped to have a minimum signal-to-noise of 6 and to oversample the energy resolution by a factor of 3. The FPM spectra are modelled over 3--60\,keV.

\subsection{\swift}

\subsubsection{XRT}
All of our X-ray Telescope (XRT) observations were operated in the photon counting mode. The calibration file version used is v.20180103. The source spectrum is extracted from a circular region with a radius of 20 pixels and the background spectrum is extracted from an annular region centered at the source with an inner radius of 40 pixels and an outer radius of 100 pixels. The spectrum is binned to have a minimum signal-to-noise of 3 and oversample by a factor of 3. The XRT spectra are modelled over 0.5--7\,keV. \blue{During obs 11 when the X-ray flux of \src\ is the highest, XRT has an averaged count rate of 0.13\,cts\,s$^{-1}$, which is much lower than the pile-up threshold of XRT\footnote{https://www.swift.ac.uk/analysis/xrt/pileup.php} (0.5\,cts\,s$^{-1}$ for the Photon Counting mode).}

\subsubsection{UVOT}

\begin{figure}
%\vspace{-2cm}
\centering
\includegraphics[width=9cm,angle=0]{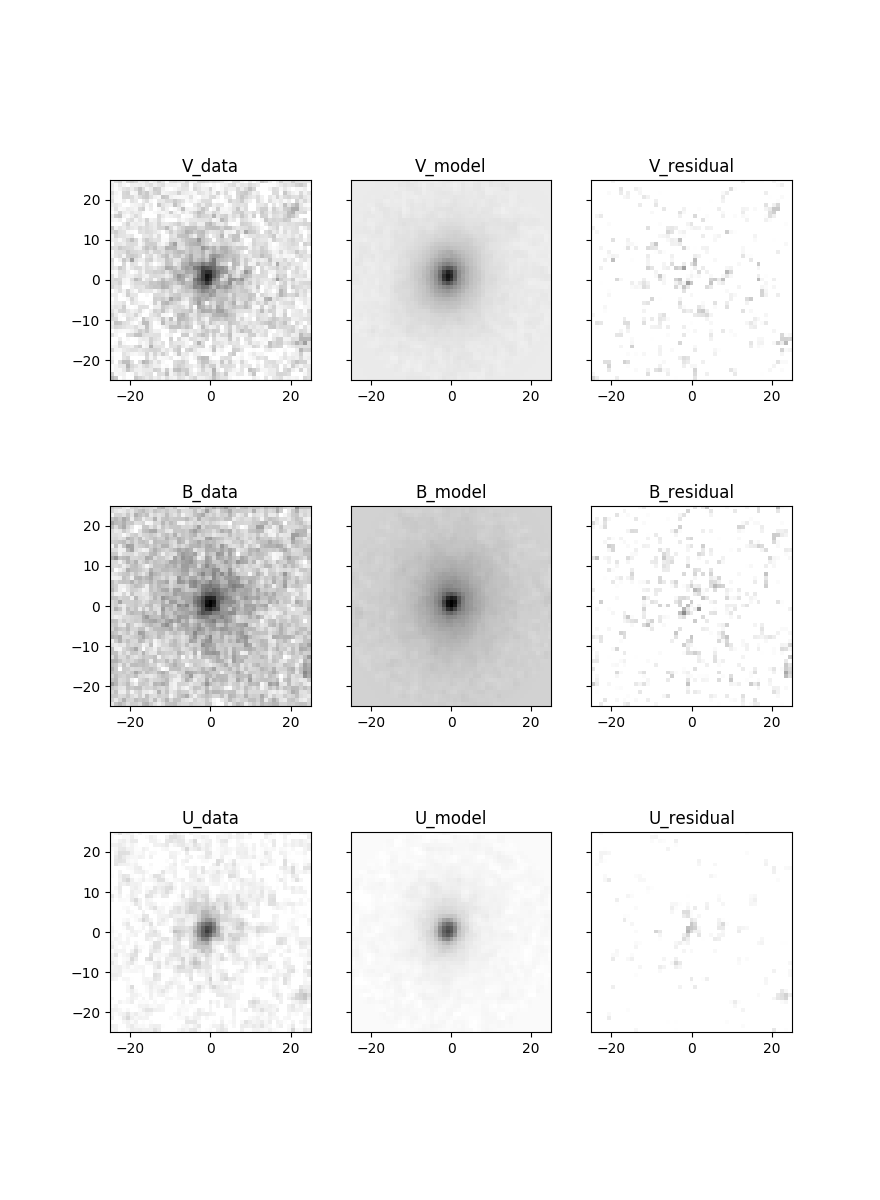}
% \begin{minipage}[b]{0.5\linewidth}
% \centering
% \includegraphics[width=3.8in,angle=0]{93060002.png}
% \end{minipage}
\vspace{-1cm}
\caption{\blue{An example of the 2-D image decomposition procedure (obs 4). The rows from top to bottom represent the results in V, B, and U bands, respectively. In each row, the first column shows the data image, the second column shows the fitting model (AGN + `exponential-disc' + background), and the third column shows the residual image derived by subtracting the model from the original data.}}
\label{fig:galfit_example}
\end{figure}

\blue{The UV and Optical Telescope (UVOT) was not operated during obs 1–3. For the rest of the {\it Swift} observations, we follow the UVOT reduction threads for the data reduction\footnote{\url{http://www.swift.ac.uk/analysis/uvot/index.php}}. Available observations with all six filters are considered. For each filter, the `Level 2' \textsc{fits} files are summed by using the \textsc{UVOTIMSUM} tool, and the source magnitudes are extracted by using the \textsc{UVOTSOURCE} tool. A circular region with a radius of 5 arcsec is used for the extraction of the source flux and a nearby circular region with a radius of 60 arcsec is used to estimate the background flux. For the three UV filters (UVW1, UVM2, and UVW2) of which the effective wavelengths are less than $\sim3000$ \AA, we simply adopt the 5 arcsec aperture photometric measurements as the AGN fluxes, as the host galaxy starlight is generally negligible within the aperture \citep[see Figures 8 and 10 in][for illustration]{add09}. While for the three optical filters (V, B, and U), the host galaxy starlight may contribute a significant fraction to the measured fluxes, and thus needs to be subtracted.}

\blue{A useful way to measure galaxies in digital images is to model their light distribution by performing two-dimensional profile fitting. \textsc{galfit} is a popular computer algorithm in which the parameters in one or more functions can be adjusted to try and match the shape and profile of galaxies \citep{add02,add10}. In this work, we implement the image decomposition on UVOT images with \textsc{galfit}. A similar procedure was also carried out in \citet{add09} and \citet{cheng19}. For each of the image, a point spread function (PSF) is constructed from nearby stars in the field of view with count rates comparable to that of KUG 1141 (within $\sim0.1$--$0.3$ dex). The background is calculated from a source-free region near the source independently and fixed in the fitting. The AGN component is modelled with the point-like source, i.e. a PSF function, with an initial magnitude from the the results of \textsc{UVOTSOURCE}. The host galaxy component is modelled with an exponential-disc function or a more general S\'ersic function, whichever shows a better fit. A more sophisticated model with additional components being included, such as a bulge or a bar can hardly be performed on UVOT data due to the low spatial resolution (the typical FWHM of the PSF is $\sim2''$). The goodness of fit is judged by both the value of the $\chi^2/\nu$, and the distribution of the residuals. In practice, it is found that for some of the observational data, the more general S\'ersic function does not work (mainly due to the low spatial resolution of the UVOT image), while the exponential-disc function appears to work throughout. Therefore, to maintain consistency, the `PSF + exponential-disc + background' model is adopted for all observations. }

\blue{We implement the image decomposition in two steps. First, for each of the 24 images (3 optical bands $\times$ 8 UVOT observations), the fitting procedure is carried out in a quick and simple way that all the parameters in the PSF and exponential-disc functions can be adjusted. It is found that for a given band, most of the disc parameters obtained in different observations, including the magnitude, effective radius, and axis ratio, are broadly consistent. For instance, the resulting magnitudes of the host galaxy remain consistent statistically over various epochs. Specifically, the fluctuations in the measured galactic flux are $\sim12$ per cent in V band, $\sim16$ per cent in B band, and $\sim15$ per cent in U band, respectively. This is fairly reasonable as the emission of the host galaxy is generally stable within the observational interval ($\sim10$ years). Next, we would like to improve the fits by taking advantage of these consistent measurements. For the image decomposition within a given band, the parameters associated with the host galaxy component can be determined by averaging the individual measurements obtained in various observations and fixed in the fitting procedure. This time only one parameter, i.e. the AGN magnitude in the PSF function, is freely fitted. In this way we conduct the image decomposition again for the eight UVOT observations in the V, B, and U bands respectively, and obtain the AGN magnitudes with the host galaxy contamination reasonably eliminated. Figure \ref{fig:galfit_example} shows an example of the image fitting, corresponding to obs 4.}

\blue{Finally, with the PSF magnitudes obtained from \textsc{galfit}, we correct for the coincidence loss effect (the phenomenon when multiple photons arrive at a similar location on the detector during a single frame, similar to the `pile-up' effect in X-rays) by Equations 1--4 in \citet{add08}, and obtain the final AGN fluxes for spectral fitting in the three optical filters. The uncertainties in these measurements are estimated by the `error propagation formula', making use of photometric errors of the whole galaxy and the fluctuations in the starlight of the galactic disc measured in multiple observations (see Appendix \ref{sec:uvoterror} for a detailed interpretation on the estimation of the uncertainties). The response file is downloaded from the {\it Swift} website\footnote{ftp://xmm.esac.esa.int/pub/ccf/constituents/extras/responses/OM} to include UVOT data in XSPEC for SED modelling.}

%The UV and Optical Telescope (UVOT) was not operated during obs 1--3. For the rest of the \swift\ observations, UVOT photometry data are extracted by using the UVOTIMSUM tool. Available observations with all six filters are considered. A circular region with a radius of 13\,arcsec is used for the extraction of the source flux. A nearby circular region with a radius of 60\,arcsec is used to estimate the background flux. We use the UVOT2PHA tool to convert the photometry data to spectra files that can be read into XSPEC. The response file is downloaded from the \swift\ website \footnote{https://swift.gsfc.nasa.gov/proposals/swift\_responses.html}.

\begin{table}
    \centering
    \begin{tabular}{cccccc}
    \hline\hline
    No. & Mission$^{\rm a}$ & Obs ID & Date$^{\rm b}$ & XRT & UVOT \\
            &  &      &      & /EPIC & /OM (ks)\\
            &  &      &      & /FPM (ks) & \\
    \hline
    1& SW& 00037136002 & 07-10-17 & 7.6 & 0 \\
    2& SW& 00037136003 & 07-10-19 & 8.9 & 0 \\
    3& SW& 00037136004 & 07-10-21 & 7.1 & 0 \\
    4& SW& 00037565001 & 09-02-21 & 1.5 & 1.4 \\
    5& XMM& 0601780501 & 09-05-23 & 2 (7, 7)$^{\rm c}$ & 16 \\
    6& SW& 00037565002 & 12-10-10 & 1.6 & 1.6 \\
    7& SW& 00037565003 & 12-10-11 & 1.6 & 1.6 \\
    8& SW& 00091632001 & 13-10-10 & 2.7 & 2.6 \\
    9& SW& 00091632002 & 13-10-11 & 2.5 & 2.4 \\
    10& SW& 00037565005 & 14-10-10 & 1.7 & 1.7 \\
    11& SW& 00093060002 & 17-11-19 & 1.6 & 1.5 \\
    12& SW& 00081097001 & 19-12-26 & 6.5 & 6.5 \\
      & Nu& 60160449002 & 19-12-26 & 21 & - \\
    \hline\hline
    \end{tabular}
    \caption{Observation details of \src. $^{\rm a}$ SW: \swift; XMM: \xmm; Nu: \nustar. $^{\rm b}$ The dates are reported in the year-month-day format. $^{\rm c}$ The net exposure time of the EPIC-pn observation after removing the time intervals when the flaring particle background dominates. The values in the bracket show the net exposure of two EPIC-MOS observations.}
    \label{obs_log}
\end{table}

\section{Long-Term Flux Variability} \label{variability}

In this section, we give an overview of the long-term variability of \src\ in the X-ray, UV and optical bands. Detailed flux values can be found in Table\,\ref{tab_uvot} and Table\,\ref{tab_flux}. 
\begin{figure}
    \centering
    \includegraphics[width=8cm]{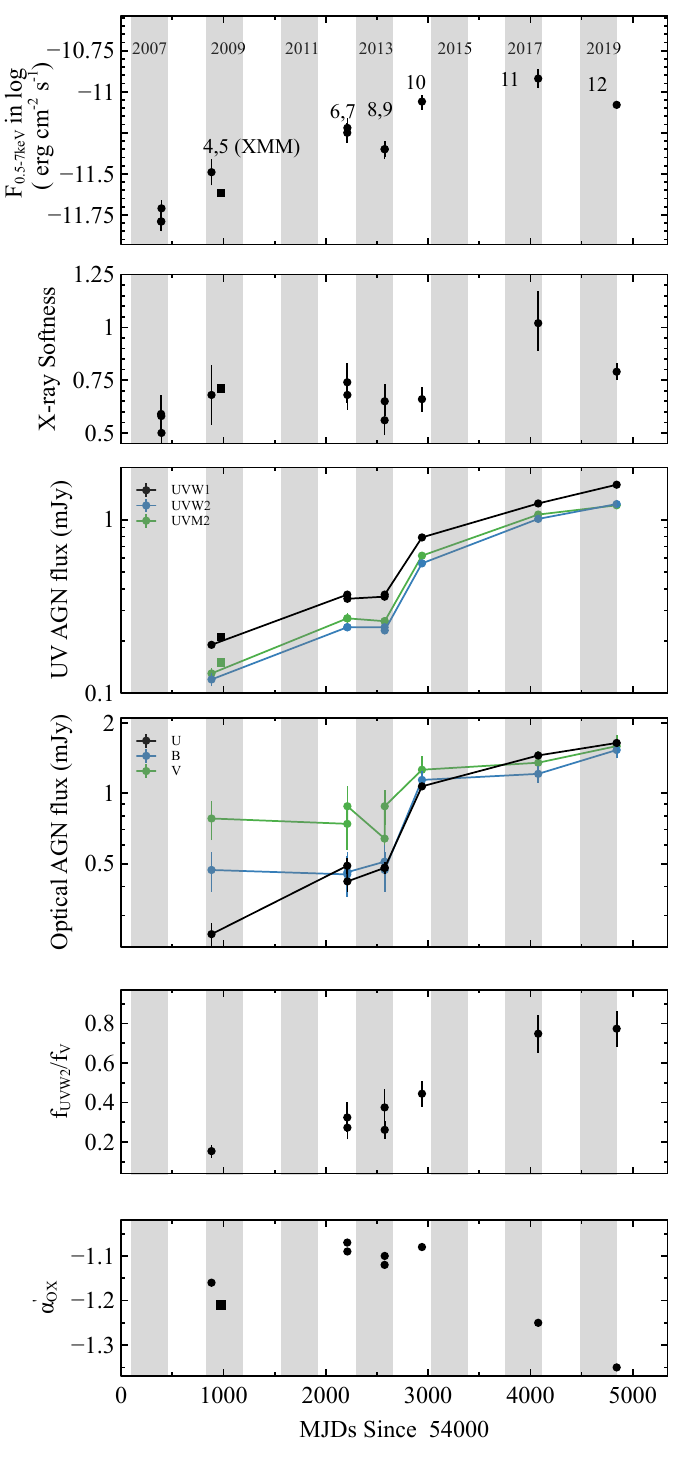}
    \caption{From top to bottom: 1) a soft X-ray (0.5--7\,keV) lightcurve of \src in units of \ergps. 2) X-ray softness defined by the flux ratio between the 0.5--2\,keV and 2--7\,keV bands. 3) AGN UV flux in units of mJy given by \swift\ UVOT and \xmm\ OM observations. 4) AGN optical flux in units of mJy given by UVOT observations. 5) Optical V band and UVW2 band flux ratio. 6) Flux density ratio between the UVW1 and 2\,keV bands. The grey bands mark calendar years.}
    \label{pic_lc}
\end{figure}

\subsection{X-rays}
A long-term X-ray lightcurve of \src\ is shown in the top panel of Fig.\,\ref{pic_lc}. The observed flux is calculated in the overlapping 0.5--7\,keV band of \swift\ XRT and \xmm\ EPIC-pn. As shown in the figure, the X-ray flux from \src\ increases by almost one order of magnitude from $\log(F_{\rm X}/$\ergps)=-11.79 in 2007 to $\log F_{\rm X}=-10.92$ in 2017 (hereafter the X-ray flux values are reported in units of \ergps). The latest observation in 2019 (obs 12) shows that \src\ still remains in a high flux state with $\log F_{\rm X}=-11.08$ since 2017.

The X-ray softness of \src, which is defined as the flux ratio between the 0.5--2\,keV and 2--7\,keV bands, is shown in the second panel of Fig.\,\ref{pic_lc}. The variability of the softness ratio indicates that the X-ray spectrum of \src\ shows a `softer-when-brighter' pattern, and is the softest in 2017 when the X-ray flux reaches the highest level. Obs 12 in 2019 shows a slightly harder continuum compared to obs 11.

\subsection{UV and optical bands}

Long-term UV and optical lightcurves \blue{of the AGN} in \src\ are shown in the third and fourth panels of Fig.\,\ref{pic_lc}. Note that the \xmm\ OM was operated with only UVW1 and UVM2 filters during obs 5. The rest of the flux values are given by the \swift\ UVOT observations in the archive. The following conclusions can be drawn:

\blue{First, the UV and optical emission from the AGN of \src\ shows simultaneous flux increase in the period of 2009--2019. For instance, UVW2 band flux has increased by more than one order of magnitude from 0.12\,mJy in 2009 to 1.23\,mJy in 2019; B band flux has increased by a factor of 3 from 0.47\,mJy in 2009 to 1.53\,mJy in 2019.}

Second, we find that the UV flux increases by a larger factor than the optical flux. In the \blue{fifth} panel of Fig.\,\ref{pic_lc}, we show of the UVW2 and V flux ratio as an example. Note that the V and UVW2 bands are respectively the longest and the shortest UVOT wavelengths. The flux ratio between these two bands indicates the steepness of the UV--optical continuum emission. As shown in Fig.\,\ref{pic_lc}, the UV--optical flux ratio is the highest in 2009 when the source is in a low optical/UV flux state; the ratio is the lowest in 2017 and 2019 when the source is in a high optical/UV flux state. 

Third, we follow the same approach as in \citet{vignali03} to calculate $\alpha_{\rm OX}$, which is defined as $0.384\log(f_{\rm 2keV}/f_{2500})$. $f_{\rm 2keV}$ and $f_{2500}$ are respectively the flux density in units of erg\,cm$^{-2}$\,s$^{-1}$\,Hz$^{-1}$ at 2\,keV and 2500\,\AA\ in the rest frame. 2500\,\AA\ in the rest frame of \src\ corresponds to 2598\,\AA\ in the observer's frame, which is within the wavelength range of the UVW1 filter (2600\,\AA) on UVOT. By approximating the UVW1 flux as the 2500\,\AA\ flux, we calculate $\alpha_{\rm OX}\approx\alpha^{\prime}_{\rm OX}=0.384\log({f_{\rm 2keV}/f_{\rm UVW1}})$. The values of $\alpha^{\prime}_{\rm OX}$ are shown by the right y-axis of the sixth panel in Fig.\,\ref{pic_lc}. $\alpha^{\prime}_{\rm OX}$ is relatively constant before 2015, and the source becomes X-ray-weak since 2017. The relative weakness of X-ray emission in \src\ might be related to enhanced reflection or absorption \citep[e.g.][]{gallo06}.

\section{X-ray Spectral Analysis} \label{xray}
\begin{figure*}
    \centering
    \includegraphics[width=18cm]{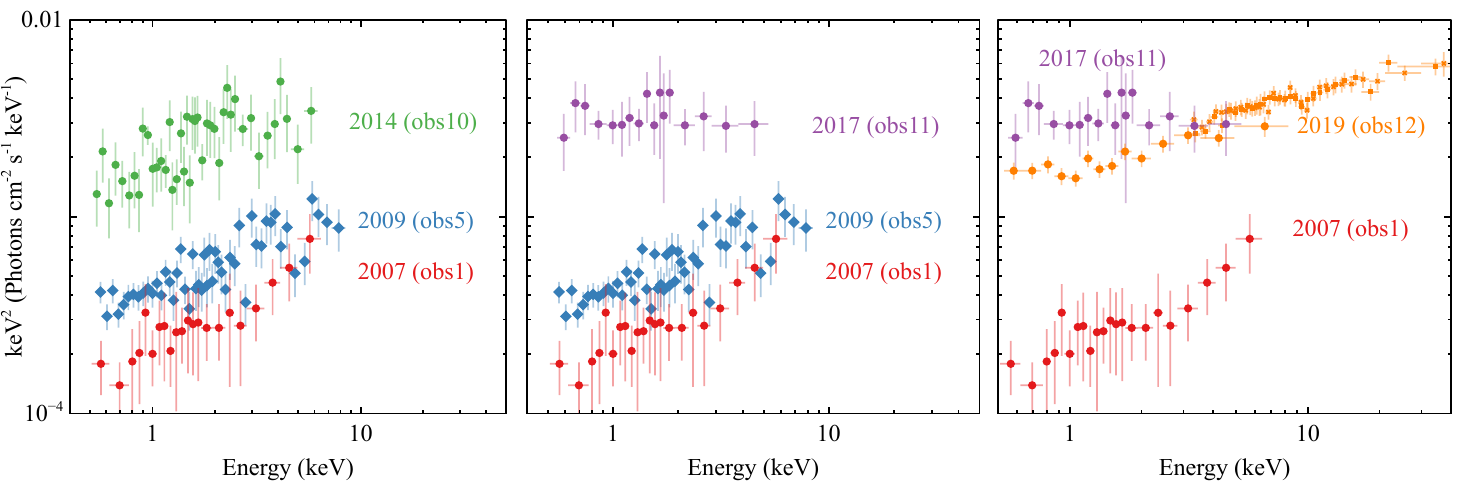}
    \caption{X-ray spectra of \src. Only some observations are shown in this figure for clarity. A power-law model with $\Gamma=0$ is used to unfold the spectra in order to remove the effects of instrumental response. Only the EPIC-pn spectrum of obs 5 (blue diamonds) is shown in the left and middle panel for simplicity. The yellow crosses and squares in the right panel show respectively the FPMA and FPMB spectra of obs 12. The rest of the spectra are all extracted from XRT observations.}
    \label{pic_eeuf}
\end{figure*}

In this section, we conduct X-ray spectral analysis for \src. Before detailed modelling, we show X-ray spectra extracted from some of the observations in Fig.\,\ref{pic_eeuf} to give an overview of the X-ray spectral variability in \src. The spectra are all unfolded using a power law with $\Gamma=0$ to remove the effects of instrumental response for demonstration purposes. 

As shown in Fig.\,\ref{pic_eeuf}, the X-ray continuum of \src\ shows a power-law shape with a steady increase of flux. In 2017, the X-ray flux during obs 11 reaches the highest level, and the X-ray continuum is the softest. For reference, a power law with $\Gamma=2$ would be a horizontal line in this figure. The latest observation (obs 12) in 2019 shows a slightly lower X-ray flux state than obs 11 with a harder X-ray continuum. The variability of the spectra agrees with the flux and X-ray softness variation shown in Fig.\,\ref{pic_lc}.

Detailed X-ray spectral analysis and SED modelling in the later section are all conducted in XSPEC \citep[v.12.10.1f,][]{arnaud96}. The line-of-sight Galactic absorption towards \src\ is $N_{\rm H}=1.76\times10^{20}$\,cm$^{-2}$, and the Galactic extinction is $E(B-V)=0.02$ \citep{willingale13}. The \texttt{tbnew} \citep{wilms00} and the \texttt{zdust} \citep{pei92} models in XSPEC are used for the Galactic X-ray absorption and optical/UV extinction. Their values are all fixed during the spectral fitting. A simple convolution model \texttt{zmshift} in XSPEC is used to account for the source redshift (z=0.038).

In the rest of this section, we start with analysing the \xmm\ (obs 5) and the \nustar\ observations (obs 12) of \src, which have higher signal-to-noise than \swift\ XRT observations. A simultaneous \swift\ snapshot observation is considered for obs 12. In the end of this section, we present the analysis of all the \swift\ XRT observations.

\begin{table}
    \caption{The best-fit values for the X-ray spectra of obs 5 (EPIC), 12 (FPM and XRT) and 11 (XRT). Model 1 is a power-law model with a high energy cutoff. Model 2 has an additional \texttt{bbody} component to test for soft excess emission. See text for more details.}
    \label{tab_small_fit}
    \centering
    \begin{tabular}{ccccccccc}
    \hline\hline
    Obs No. & Parameter & Unit & Model 1 & Model 2 \\
    \hline
    5 & $\Gamma$ & -   & $1.71\pm0.04$ & $1.60\pm0.10$\\
    \textit{XMM}  & Ecut & keV & 500 & 500 \\
      & $kT$     & keV & -             & $<0.25$ \\
      & norm     & $10^{-5}$ & - & unconstrained \\
      & $\chi^{2}/\nu$ & - & 155.12/164 & 149.03/162\\
    \hline
    12 & $\Gamma$ & - & $1.71\pm0.03$ & $1.67^{+0.03}_{-0.08}$\\
    \nustar   & Ecut & keV & >220 & >90 \\
    \& \swift   & kT & keV & - & $0.10^{+0.04}_{-0.03}$ \\
       & norm & $10^{-5}$ & - & $3^{+9}_{-2}$ \\
       & $\chi^{2}/\nu$ & - & 427.43/402 & 413.16/400\\
    \hline
    1 & $\Gamma$ & - & $1.54\pm0.19$ & $1.54\pm0.19$\\
    \swift   & Ecut & keV & 500 & 500 \\
       & kT & keV & - & 0.10 \\
       & norm & $10^{-5}$ & - & $<0.4$ \\
       & $\chi^{2}/\nu$ & - & 20.69/20 & 20.60/19\\
    \hline
    11 & $\Gamma$ & - & $2.0\pm0.2$ & $1.9\pm0.3$\\
    \swift   & Ecut & keV & 500 & 500 \\
       & kT & keV & - & 0.10 \\
       & norm & $10^{-5}$ & - & $<10$ \\
       & $\chi^{2}/\nu$ & - & 27.88/15 & 27.60/14\\
    \hline\hline
    \end{tabular}
\end{table}

\subsection{\xmm\ (obs 5)} \label{obs5}

\begin{figure*}
    \includegraphics[width=8cm]{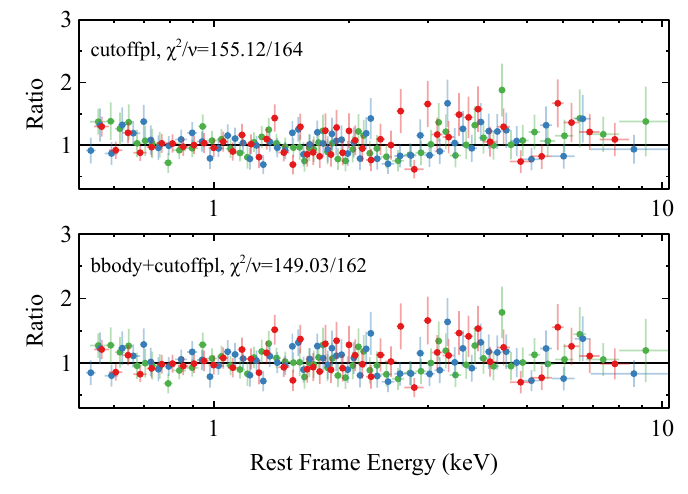}
    \includegraphics[width=9cm]{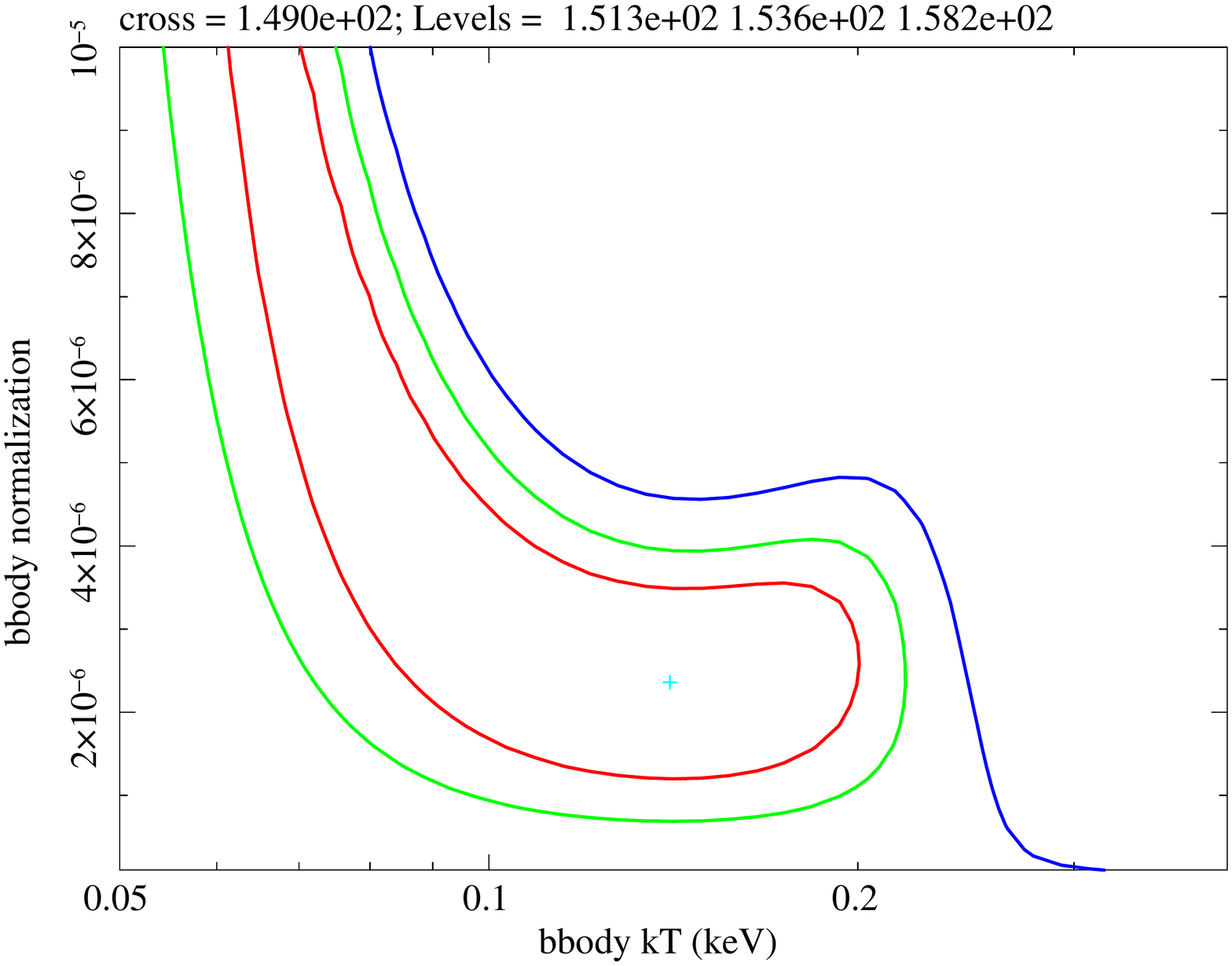}
    \caption{Left: ratio plots for the EPIC spectra of obs 5 using different continuum models. Red: pn; blue: MOS1; green: MOS2. An additional \texttt{bbody} component improves the fit by $\Delta\chi^{2}=6$ with 2 more free parameters. Right: a contour plot of $\chi^{2}$ distribution on the plane of $kT$  vs. the normalization of the \texttt{bbody} model.  The red, green and blue contours show the 1, 2, 3 $\sigma$ regions.}
    \label{pic_obs5}
\end{figure*}

We first model the three EPIC spectra of obs 5 with an absorbed power-law model \texttt{cutoffpl} (Model 1). Due to the lack of simultaneous hard X-ray observation, we fix the high energy cutoff parameter $E_{\rm cut}$ at 500\,keV. Model 1 is able to fit the EPIC spectra very well with $\chi^{2}_{\rm red}=0.95$. The best-fit parameters are shown in Table\,\ref{tab_small_fit}, and the corresponding ratio plot is shown in Fig.\,\ref{pic_obs5}. \blue{We did not find obvious evidence for narrow emission feature or absorption edge features in the iron band.}

In order to test for possible soft excess emission, we add an additional \texttt{bbody} model (Model 2), which improves the fit by only $\Delta\chi^{2}=6$ with 2 more parameters. See the left panel of Fig.\,\ref{pic_obs5} for comparison of the fits. Only an upper limit of the temperature parameter is obtained ($kT$<0.2\,keV). The normalization parameter of the \texttt{bbody} model is not constrained. A contour plot of $\chi^{2}$ distribution on the $kT$ vs. normalization parameter plane is shown in the right panel of Fig.\,\ref{pic_obs5}. The normalization of the \texttt{bbody} component is consistent with a very low value, e.g. <$1\times10^{-6}$, within a 3$\sigma$ uncertainty range. When $kT$ is at a very low value, e.g. <0.07\,keV, the fit is no longer sensitive to $kT$ due to the lower limit of the energy coverage of EPIC.

We also test for any extra absorption along the line of the sight towards \src\ by adding an additional \texttt{tbnew} model at the source redshift. We only obtain an upper limit of the column density (a 90\% confidence range of $N_{\rm H}<3\times10^{20}$\,cm$^{-2}$). 

To sum up, we conclude that the X-ray spectra of \src\ during obs 5 are consistent with a Galactic-absorbed power law. There is no/little evidence for soft excess emission or additional line-of-sight absorption.

\subsection{\nustar\ (obs 12)} \label{obs12}
\begin{figure*}
    \includegraphics[width=8cm]{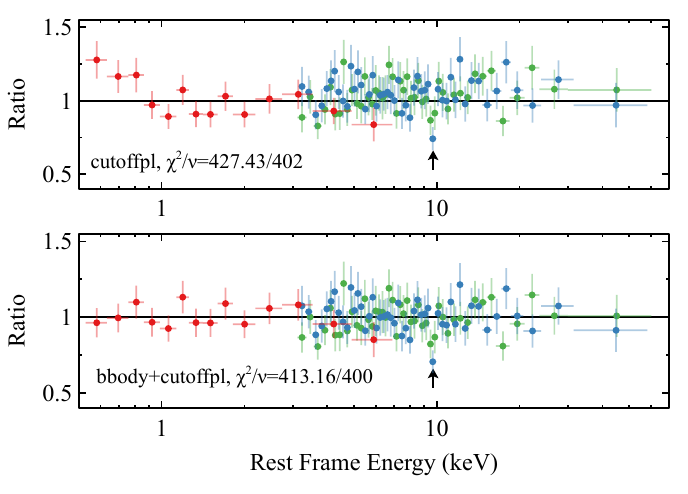}
    \includegraphics[width=9cm]{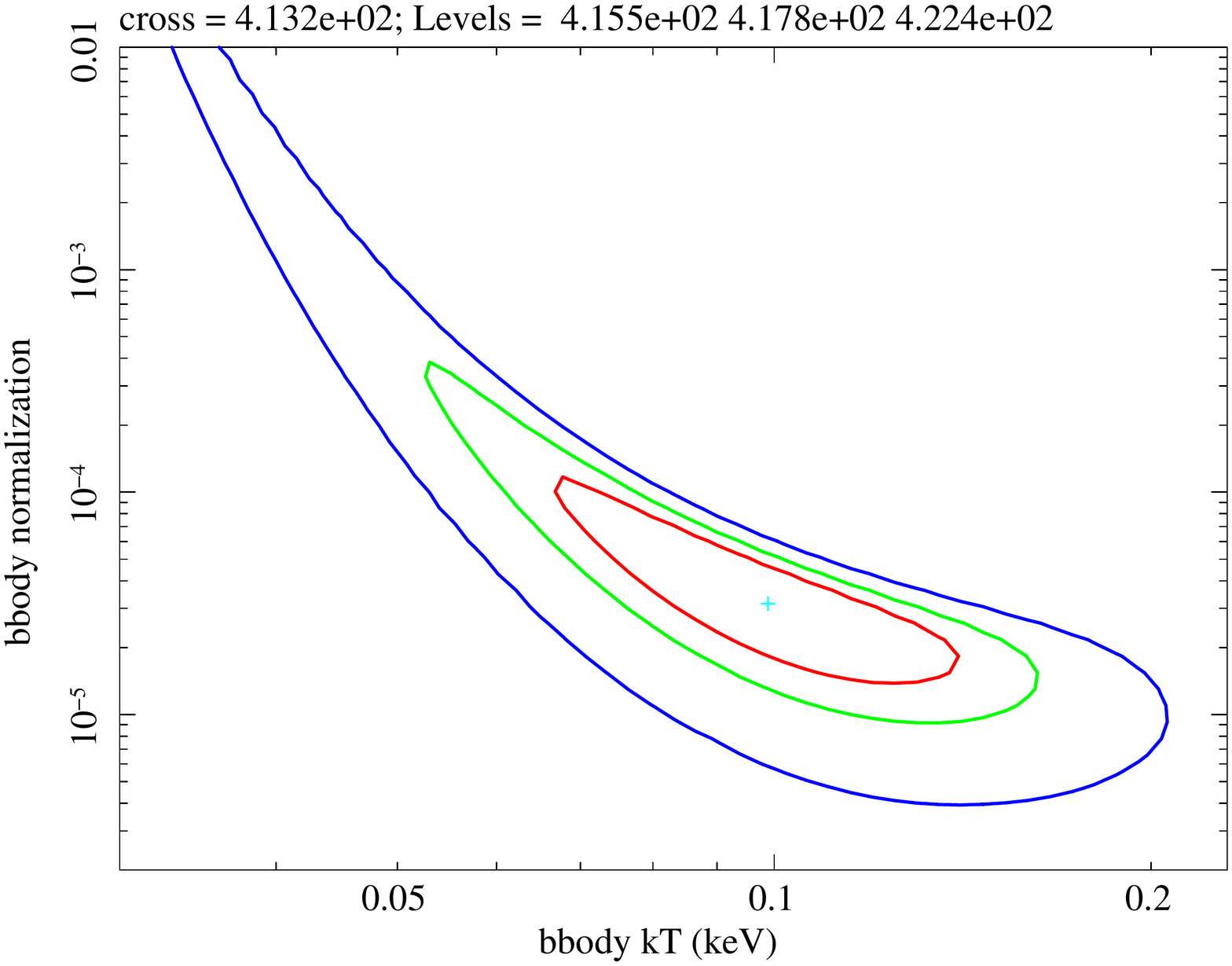}
    \caption{Left: ratio plots for the FPM and XRT spectra of obs 12 using different continuum models. Red: XRT; blue: FPMA; green: FPMB. An additional \texttt{bbody} component improves the fit by $\Delta\chi^{2}=14$ with 2 more free parameters. The black arrows show the narrow absorption line in the FPMA and FPMB spectra. Right: a contour plot of $\chi^{2}$ distribution on the plane of $kT$  vs. the normalization of the \texttt{bbody} model.  The red, green and blue contours show the 1, 2, 3 $\sigma$ regions.}
    \label{pic_obs12}
\end{figure*}

\subsubsection{Continuum emission} \label{obs12_con}
We first follow the approach in Section\,\ref{obs5} by modelling the FPM and simultaneous XRT spectra of obs 12 with \texttt{cutoffpl} (Model 1). The best-fit parameters are shown in Table\,\ref{tab_small_fit}. A power law with a high energy cutoff of $E_{\rm cut}$>220\,keV is able to describe the spectra very well above 1\,keV. \blue{Similar to \xmm\ spectra, we did not find obvious evidence for any narrow emission line or absorption edge in the iron band of FPM spectra. Therefore, we conclude that the spectra have no or little contribution from any distant cold reflector.}

However, we find clear evidence for soft excess emission below 1\,keV (see the left top panel of Fig.\,\ref{pic_obs12}). An additional \texttt{bbody} model for the soft excess emission (Model 2) can improve the fit by $\Delta\chi^{2}=14$ with 2 more free parameters. The constraint of the parameters of the \texttt{bbody} component is shown in the right panel of Fig.\,\ref{pic_obs12}. The best-fit $kT$ is $0.10^{+0.04}_{-0.08}$\,keV, which is similar to the typical value for other Sy1s \citep[e.g.][]{walter93}. Such soft excess emission is however not found in the \xmm\ observation (obs 5) in 2009. Note that Model 2 requires a slightly harder continuum ($\Gamma=1.67$) with a smaller lower limit of the high energy cutoff ($E_{\rm cut}$>90\,keV) than Model 1. 

\subsubsection{Ultra-fast outflow?}
We also notice that there is evidence for a weak narrow absorption line between 9--10\,keV in both the FPMA and FPMB spectra of obs 12. The black arrows in the left panels of Fig.\,\ref{pic_obs12} show the position of the line. This feature is often interpreted as a blueshifted Fe~\textsc{xxv} or Fe~\textsc{xxvi} line from an ultra-fast outflow \citep[e.g.][]{tombesi11}. By modelling the absorption line feature with the \texttt{zgauss} model, we obtain the best-fit line energy of $9.72\pm0.01$\,keV at the source frame, line width of $\sigma<0.28$\,keV, equivalent width of $136\pm20$\,eV. This additional line model is able to improve the fit by $\Delta\chi^{2}=12$ with 3 more parameters, corresponding to a F-statistic value of 4.3 with probability of 0.005.
\begin{figure}
    \centering
    \includegraphics[width=8cm]{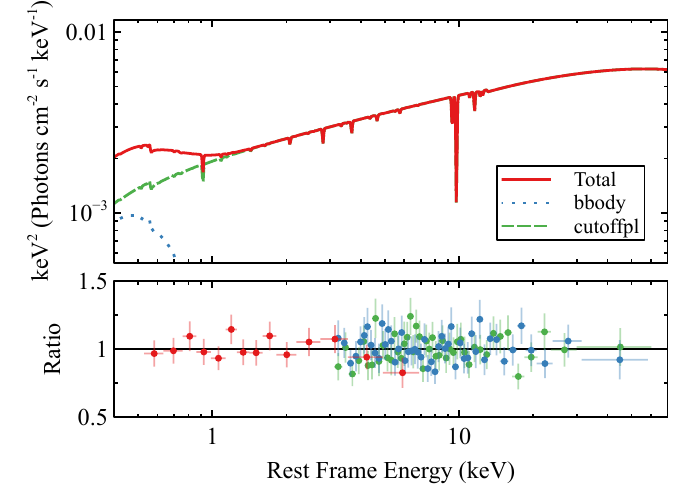}
    \caption{Top: the best-fit model for obs 12 after considering absorption from an ultra-fast outflow. Red solid line: total model; green dashed line: power-law continuum; blue dotted line: blackbody component. Bottom: the corresponding data/model ratio plot. Red: XRT; blue: FPMA; green: FPMB.}
    \label{pic_xstar}
\end{figure}

We construct a photo-ionisation absorption grid model by using \texttt{xstar} \citep{kallman01} assuming a power-law illuminating spectrum with $\Gamma=1.7$. Solar abundances are assumed. A turbulent velocity of 2000 km\,s$^{-1}$ is used. The free parameters are the column density ($N_{\rm H}$), the ionisation state of the absorber ($\xi$), and the blueshift parameter ($z$). By modelling the absorption line with \texttt{xstar}, we obtain $N_{\rm H}=(4\pm3)\times10^{23}$\,cm$^{-2}$, $\log(\xi)=3.7\pm0.4$ and $z=0.258^{+0.011}_{-0.015}$. The value of $z$ is calculated in the observer's frame, which corresponds to a line-of-sight velocity of $v=0.32\pm0.02c$.  

According to the best-fit \texttt{xstar} model, the absorption line at 9.72\,keV is mainly Fe~\textsc{xxvi} line with a little contribution from the Fe \textsc{xxv} line. See Fig.\,\ref{pic_xstar} for the best-fit absorption model. FPMs are unable to resolve these two lines due to their limited energy resolution. There was no simultaneous high resolution observation in soft X-rays, e.g. from \xmm\ RGS or \chandra\ LEGT. Thus, we are also unable to check for blueshifted O~\textsc{vii} line as predicted by our model. To sum up, we only point out the tentative evidence for an ultra-fast outflow with line-of-sight velocity of $v\approx0.32c$ during obs 12, which was taken when \src\ is in a high X-ray and optical/UV flux state.

\subsection{\swift}
\begin{figure*}
    \includegraphics[width=18cm]{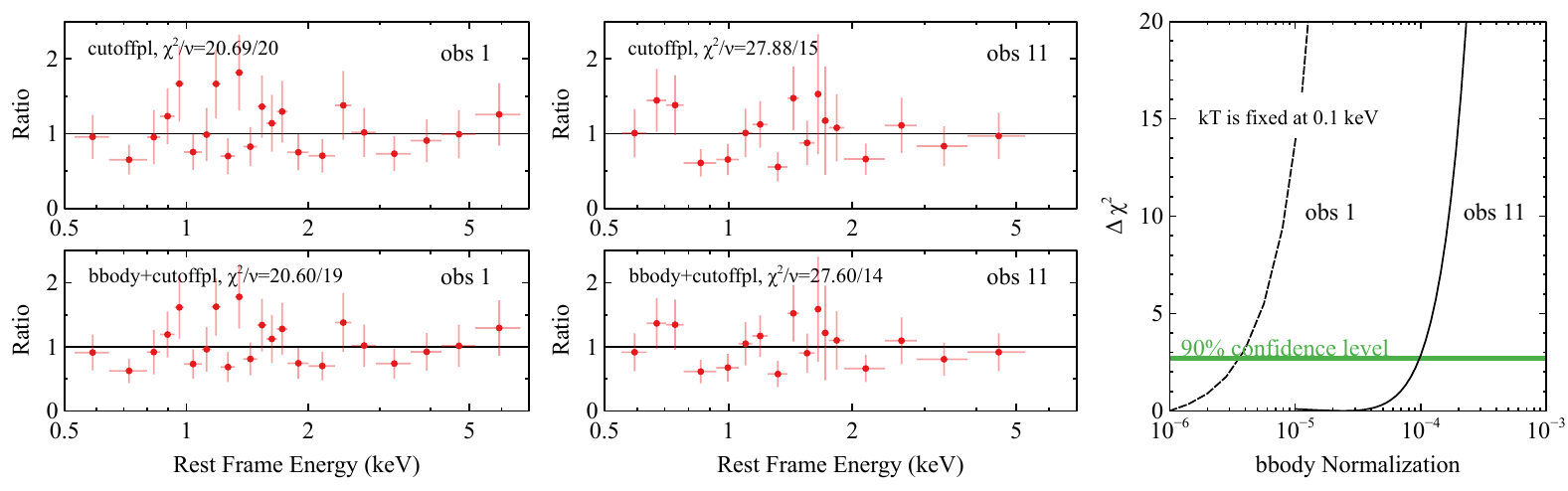}
    \caption{Left two panels: ratio plots of the XRT spectra of \src\ extracted from obs 1 and 11, which show the lowest flux state and the highest flux state in the \swift\ archive, using different continuum models. An additional \texttt{bbody} component does not improve the fit significantly. Right: the constraint of the normalization parameter of the \texttt{bbody} component with $kT$ fixed at 0.1\,keV.}
    \label{pic_obs1}
\end{figure*}

\subsubsection{Two extreme flux states: Obs 1 and 11}
Obs 1 and obs 11 were taken by \swift\ in 2007 and 2017. During these two observations, \src\ shows respectively the historical lowest and the highest X-ray flux states in the \swift\ archive. 

We first model the XRT spectra of these two observations with \texttt{cutoffpl} with $E_{\rm cut}$ fixed at 500\,keV. The best-fit values are shown in Table\,\ref{tab_small_fit}, and the corresponding ratio plots are shown in the left two panels of Fig.\,\ref{pic_obs1}. The absorbed power-law models are able to describe the data very well with $\chi^{2}/\nu=20.69/20$ for obs 1 and $\chi^{2}/\nu=27.88/15$ for obs 11. The best-fit photon indices for obs 1 and 11 are $1.54\pm0.19$ and $2.0\pm0.2$ respectively, suggesting a softer X-ray continuum in the highest flux state than in the lowest flux state. 

Alternatively, the spectral difference between obs 11 and obs 1, as shown in the middle panel of Fig.\,\ref{pic_eeuf}, might be due to not only a change in the intrinsic flux but also a variable line-of-sight column density. For example, we fit the spectrum of obs 1 with an additional neutral absorption model \texttt{tbnew} at the source redshift. The photon index of the power-law continuum is fixed at the best-fit value for obs 11 ($\Gamma=2$). An additional absorber with a column density of $(1.3\pm0.6)\times10^{21}$\,cm$^{-2}$ is then required to fit the spectrum of obs 1. Such a model also provides a good fit with $\chi^{2}/\nu=21.65/20$. In this scenario, the intrinsic X-ray continuum emission of \src\ remains similar steepness with only an increase of flux from 2007 to 2017. An additional variable Compton-thin absorber is required to explain the spectral variation. However, it is important to point out that the \xmm\ observation (obs 5), which was also taken in a low flux state, rules out the possibility of such an absorber (see Section\,\ref{obs5}). Therefore, we conclude that the scenario of variable intrinsic continuum emission is preferred rather than a Compton-thin absorber crossing the line of sight in coincidence with a change in the intrinsic X-ray flux.

Second, as shown in Fig.\,\ref{pic_obs1}, there is no significant evidence for soft excess emission. We here put an upper limit of the soft excess by assuming the best-fit $kT$ given by obs 12 ($kT=0.1$\,keV, see Section\,\ref{obs12_con}). We show $\Delta\chi^{2}$ vs. the normalization of the \texttt{bbody} model component for obs 1 and obs 11 in the right panel of Fig.\,\ref{pic_obs1}. We obtain an upper limit of $<4\times10^{-6}$ and $<1\times10^{-4}$ for obs 1 and obs 11 respectively. The inclusion of such weak components does not affect the power-law continuum modelling (see Table\,\ref{tab_small_fit} for comparison of Model 1 and 2). In summary, we find no evidence of soft excess emission in these two \swift\ observations of \src. 

\subsubsection{Other \swift\ observations}

\begin{table}
    \centering
        \caption{The best-fit Model 1 for \swift\ XRT observations of \src. The parameters for obs 1, 5, 11 and 12 can be found in Table\,\ref{tab_small_fit}.}
    \label{tab_small_fit2}
    \begin{tabular}{ccc}
    \hline\hline
    Obs No. & $\Gamma$ & $\chi^{2}/\nu$ \\ 
    \hline
    2  & $1.54\pm0.15$ & 27.92/31 \\
    3  & $1.43\pm0.16$ & 32.25/29 \\
    4   & $1.65^{+0.20}_{-0.18}$ & 4.12/9\\
    6   & $1.7\pm0.2$ & 24.32/21\\
    7   & $1.67\pm0.18$ & 15.30/23\\
    8   & $1.52\pm0.18$ & 31.63/30\\
    9   & $1.62\pm0.16$ & 32.36/30\\
    10  & $1.65\pm0.13$ & 49.59/40\\
    \hline\hline
    \end{tabular}
\end{table}

Following the conclusions above, we model the XRT spectra extracted from the other \swift\ observations using an absorbed power-law model by following the same approach. The best-fit parameters are shown in Table\,\ref{tab_small_fit2}, and the corresponding ratio plots can be found in Fig.\,\ref{pic_xrt_ratio}. Similarly, we do not find statistically significant evidence for soft excess in these observations. The best-fit photon index shows a slightly increase with the observed X-ray flux from $\approx1.5$ in 2007 to $\approx1.7$ in 2014. However, the values are statistically consistent within a 90\% confidence range. By comparing these \swift\ observations before 2015 with obs 11 in 2017, we find that the intrinsic continuum emission of \src\ is significantly softer in the highest X-ray flux state ($\Gamma=2.0\pm0.2$). 

\section{Multi-Wavelength SED Analysis} \label{sed}
So far, we have obtained the best-fit models for the X-ray continuum emission in all the epochs. We find that the X-ray spectra of obs 1-11 all show a power-law shape. Only obs 12 shows weak soft excess emission below 1\,keV, which is consistent with a blackbody with $kT=0.1$\,keV. No significant evidence of soft excess emission is found in the soft X-ray band of other epochs. No additional line-of-sight absorption was found during obs 5 (\xmm) when the source was in a low flux state and obs 12 (\nustar\ and \swift) when the source was in a high flux state. In this section, we model the multi-wavelength SEDs of \src\ in all the epochs with simultaneous observations from UVOT or OM (obs 4-12).

\subsection{SED Modelling}
\begin{figure*}
    \centering
    \includegraphics[width=18cm]{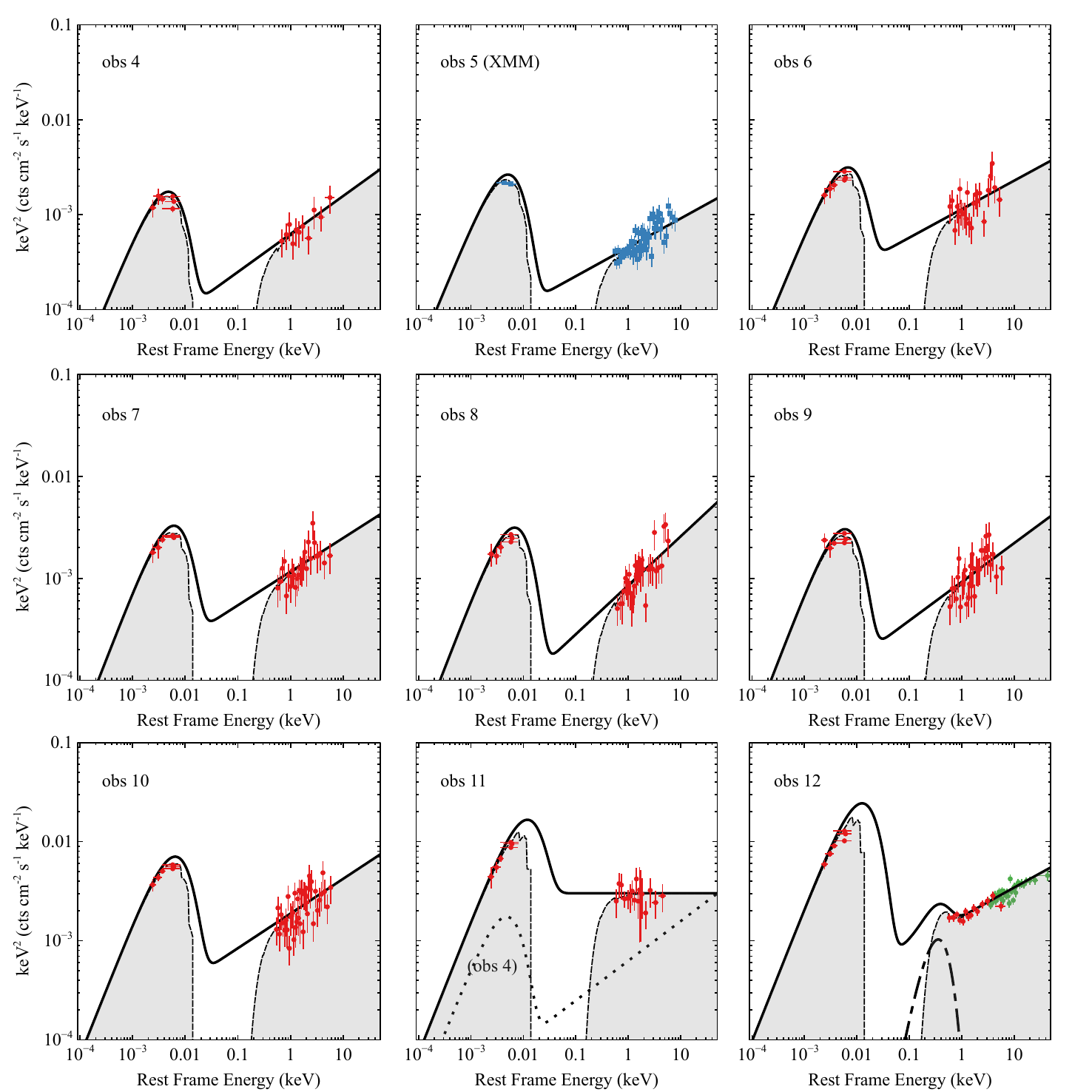}
    \caption{\blue{Multi-wavelength SEDs of \src\ extracted from all the epochs with simultaneous UV or optical observations. Red circles: \swift; blue squares: \xmm; green diamonds: \nustar. The MOS spectra of obs 5 and the FPMB spectrum of obs 12 are not shown in the corresponding panels for clarity. The dashed lines are the best-fit SED models. The black solid lines are the best-fit AGN component after removing Galactic absorption and extinction. We show the best-fit AGN model for obs 4 (dotted line) in the bottom middle panel, which is the same as the solid line in the first panel, in comparison with obs 11. The dot-dashed line in the last panel is the \texttt{bbody} component that is used to model the soft excess emission in obs 12.}}
    \label{pic_sed}
\end{figure*}

We first model the disc thermal emission of \src\ using the \texttt{diskbb} model. This model assumes a thin disc temperature profile of $T\propto r^{-0.75}$ and has two parameters: the inner temperature of the disc ($kT_{\rm in}$) and the normalization parameter. The normalization parameter is defined as $(R^{\prime}_{\rm in}/D_{\rm 10})^{2}\cos i$, where $R^{\prime}_{\rm in}$ is the `apparent' inner radius of the disc in km, $D_{\rm 10}$ is the source distance in 10\,kpc, and $i$ is the inclination angle of the disc. See Section\,\ref{sed_disc} for more discussion concerning the normalization parameter of \texttt{diskbb}. Note that both $kT_{\rm in}$ and the normalization parameter can change the total flux of the model. 

Secondly, we calculate the Comptonisation spectrum of the hot coronal region using the convolution model \texttt{simpl} \citep{steiner09}. The \texttt{simpl} model calculates a power law-shaped Comptonisation spectrum that self-consistently accounts for the fraction of scattered radiation. The spectrum of the seed photons is the \texttt{diskbb} model as described above. In this way, we are able to constrain the strength of the disc thermal emission by consistently considering the scattering process, and thus estimate the inner radius of the disc ($R_{\rm in}$). Such a method was previously used to measure $R_{\rm in}$ in BH transients when the sources are not necessarily in a thermal-dominant state \citep[e.g.][]{steiner09}. The free parameters are the scattering fraction $f_{\rm scatt}$ and the photon index of the power-law continuum ($\Gamma$). 

\begin{figure}
    \centering
    \includegraphics{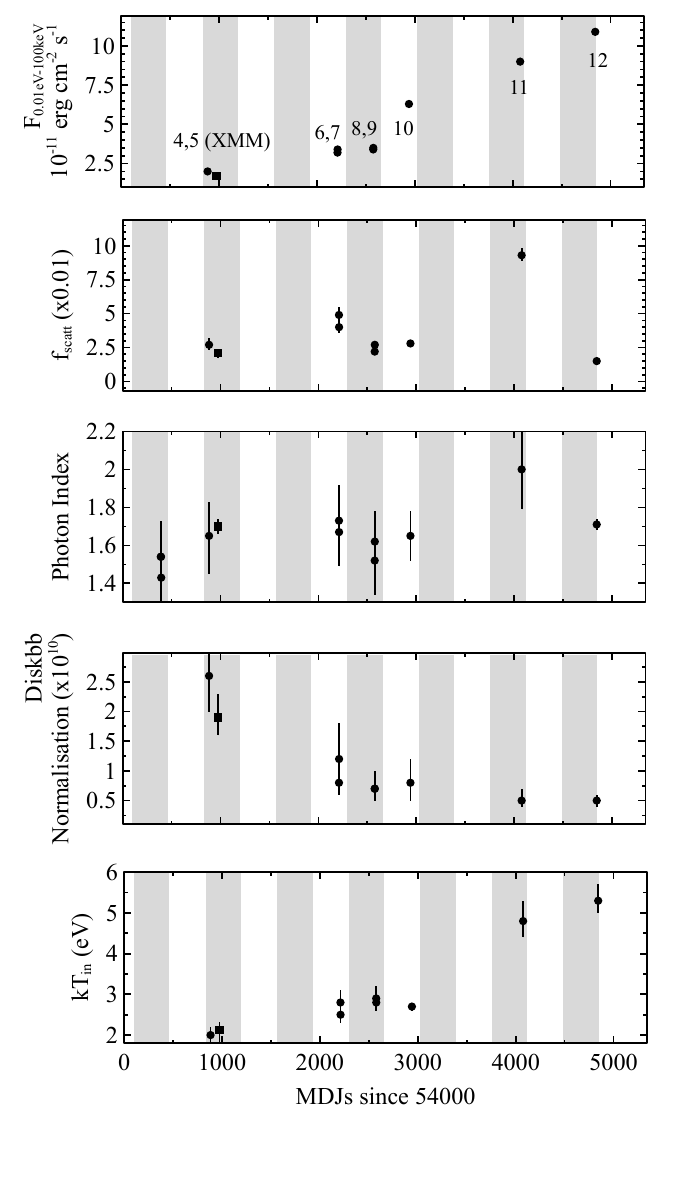}
    \caption{\blue{The best-fit SED model parameters for each epoch. Circles: \swift; sqaures: \xmm. The best-fit photon indexes for obs 1--3 are obtained from the analysis of only X-ray data.}}
    \label{pic_sed_fit}
\end{figure}

The total model is \texttt{tbnew * zdust * zmshift * ( simpl * diskbb )} in the XSPEC format. Such a model can describe the SEDs of most epochs very well except obs 12. An additional weak \texttt{bbody} model is used to model the soft excess shown in obs 12 as demonstrated in Section\,\ref{obs12_con}. The best-fit SED models are shown in Fig.\,\ref{pic_sed}, and the best-fit parameters are shown in Table\,\ref{tab_sed} and Fig.\,\ref{pic_sed_fit}. In the top panel of Fig.\,\ref{pic_sed_fit}, we also show the absorption-corrected broad band flux calculated between 0.01\,eV--100\,keV using our best-fit AGN component. The following conclusions can be drawn from our SED modelling:

1) An increase of the disc temperature together with an increase of the broad band flux is clearly seen according to our SED modelling. \blue{For instance, the best-fit inner temperature of the disc is $kT_{\rm in}=2.0\pm0.2$\,eV during obs 4; a higher temperature of $kT_{\rm in}=4.8^{+0.5}_{-0.4}$\,eV is required for obs 11.} See the bottom middle panel of Fig.\,\ref{pic_sed} for comparison of these two epochs. The increase of the disc temperature is able to explain the variability of the optical--UV continuum: the UV emission is more sensitive to the increase of the disc temperature than the optical emission, and thus increases by a larger factor than the optical emission as shown in fifth panel of Fig.\,\ref{pic_lc}.

2) \blue{Despite an increase of the disc temperature, the normalization of the \texttt{diskbb} model decreases from $\approx2.6\times10^{10}$ in 2009 to $\approx5\times10^{9}$ in 2019.} Note that, again, an increase of either $kT_{\rm in}$ or the normalization parameter can increase the flux of the model. The decrease of the normalization parameter suggests a decreasing inner radius of the disc as this parameter is proportional to $R^{2}_{\rm in}$. See Section\,\ref{sed_disc} for more discussion concerning the inner radius of the disc.

3) The scattering fraction parameter ($f_{\rm scatt}$) shows the number fraction of the disc seed photons that are up-scattered to the X-ray band in the coronal region and gives a physical interpretation of the X-ray--UV ratio shown in Fig.\,\ref{pic_lc}. $f_{\rm scatt}$ is the highest during obs 11, where the X-ray flux and the X-ray--UV ratio are both shown to be the highest among all the epochs. It is interesting to note that the best-fit $f_{\rm scatt}$ for obs 12 is similar to the values for obs 4-10 when the source is in a much lower UV and optical flux state. 

A change in  the coronal region might be a possible explanation: during the high flux state during obs 11, a different corona, e.g. of a larger size in which more disc seed photons are up-scattered, might exist. Therefore, a high fraction of accretion power is released in the form of non-thermal X-ray emission rather than the thermal emission from the disc during obs 11. The corona during obs 12 might however be similar to those during obs 4-10 despite a higher mass accretion rate. Our studies suggest that the X-ray--UV ratio/$f_{\rm scatt}$ do not necessarily show strict correlation with mass accretion rate in an individual AGN.

4) The photon index of the X-ray continuum ($\Gamma$) changes with $f_{\rm scatt}$. This suggests that the optical depth of the corona is sensitive to not only the mass accretion rate as shown by statistical studies of AGN surveys \citep[e.g.][]{brightman13} but also the energy interplay between the disc and the corona.

\subsection{Estimating $\lambda_{\rm Edd}$ and $R_{\rm in}$} \label{sed_disc}

Assuming $L_{\rm Bol}\approx L_{\rm 0.01eV-100keV}$ and a BH mass of $10^{8}M_{\odot}$\footnote{
\citet{oh15} estimated $\log(M_{\rm BH})=7.99\pm0.06$ (statistical error) for \src\ based on the study of H$\alpha$ emission line by following the method in \citet{greene05}. Systematic errors of this method can lead to a larger uncertainity range, such as the relation between $\rm FWHM_{\rm H\alpha}$ and $\rm FWHM_{\rm H\beta}$ \citep{greene05} and the uncertain geometry of the broad-line region \citep{kaspi00,mclure04}. 
}, we estimate the Eddington ratio of \src\ in the past decade. For example, \blue{a broad band flux of $F_{\rm 0.01eV-100keV}=2\times10^{-11}$\,\ergps\ during obs 4 corresponds to an Eddington ratio of $\lambda_{\rm Edd}=0.6\%$. Our best-fit SED model suggests that the bolometric luminosity of \src\ increases from $\approx 0.6\%$ of the Eddington limit in 2009 to $\approx 3.2\%$ in 2019.} A similar fraction of mass accretion rate increase has also been seen in other flaring AGN \citep[e.g. NGC~1566,][]{parker19a}. 

From the normalization parameter of the \texttt{diskbb} model, we are also able to estimate the inner radius of the disc. This parameter is defined as $(R^{\prime}_{\rm in}/D_{\rm 10})^{2}\cos i$ as introduced above, where $R^{\prime}_{\rm in}$ is the `apparent' inner radius. We assume a correction factor of 0.412 \citep{kubota98} to convert $R^{\prime}_{\rm in}$ to the real inner radius $R_{\rm in}$. For instance, \blue{assuming a viewing angle of $i=10^{\circ}$, the best-fit normalization parameter of \texttt{diskbb} for obs 4 is $2.6\times10^{10}$ corresponding to an inner radius of $R_{\rm in}\approx8$\,$r_{\rm g}$. Based on our calculations, the inner disc radius of \src\ decreases from 8\,$r_{\rm g}$ in 2009 to 3\,$r_{\rm g}$ in 2017 and 2019 assuming $i=10^{\circ}$. An assumption for a higher viewing angle leads to a larger estimated value: $R_{\rm in}$ decreases from 20\,$r_{\rm g}$ in 2009 to 8\,$r_{\rm g}$ in 2019 assuming $i=80^{\circ}$.} However, \src\ is a Sy1 galaxy which is unlikely to be viewed from an edge-on angle according to the standard model of AGN \citep[e.g.][]{antonucci93} assuming the torus plane and the optical-emitting disc plane share the same inclination.

\blue{In conclusion, we find evidence for a possible decreasing inner radius simultaneously with an increasing mass accretion rate in \src, and the value $R_{\rm in}$ has decreased by a factor of approximately 2.7.} The assumptions on viewing angle and black hole mass do not affect the change of parameters we observe but only the absolute values.

\section{Discussion} \label{discuss}

\subsection{The Behaviour of the Disc in \src}

\citet{runco16} studied the SDSS and Keck observations of \src\ in the optical band, which were taken respectively in 2005 and 2009. These optical spectra of \src\ are consistent with a typical Sy1. Meanwhile, these two observations are mostly consistent, which suggests that the disc of \src\ does not show large variability in the period of 2005-2009.

By analysing the observations after 2009, we find that \src\ has shown a steady increase of optical and UV flux. The luminosity of \src\ increases from 0.6\% of the Eddington limit in 2009 to 3.2\% in 2017. The latest UVOT observation suggests that UV and optical flux of \src\ is still increasing. 

Detailed SED modelling shows that the inner radius of the disc ($R_{\rm in}$) in \src\ decreases by a factor of approximately $2.7$. \blue{The exact value of $R_{\rm in}$ depends on the assumption for the disc inclination angle: $R_{\rm in}$ decreases, for example, from 8\,$r_{\rm g}$ (20\,$r_{\rm g}$) in 2009 to 3\,$r_{\rm g}$ (8\,$r_{\rm g}$) in 2019 assuming $\theta=10^{\circ}$ ($80^{\circ}$).} 
%The slightly truncated disc might explain the lack of evidence for apparent broad Fe~K emission line as shown in typical Sy1 \citep[e.g.][]{fabian03}. 
\citet{saxton15} estimated the filling time of a truncated thin disc in the framework of the standard thin disc model \citep{shakura73}. They found that, for example, it takes more than 800 years to refill a truncated disc with $R_{\rm in}=20$\,$r_{\rm g}$ around a $10^{8}$\,$M_{\odot}$ BH as in \src\ based on viscous timescales. Therefore, the change of $R_{\rm in}$ in \src\ is much faster than expected.

The mechanisms behind a quick boost of mass accretion rate as in \src\ are also still unclear. It might be related to the instability of a gas pressure-dominated disc at a low Eddington ratio \citep[][]{saxton15}, the propagation process in the disc \citep[e.g.][]{ross18} or tidal disruption events \citep[TDEs,][]{rees88}. 

In the case of TDEs, it is interesting to note that theoretically the rise time for a TDE to reach the peak luminosity is less than 1 year, assuming a 1~$M_{\odot}$ star disrupted by a $10^{8}$\,$M_{\odot}$ BH as in \srcfull\ and the tidal radius is twice the periastron radius \citep{decolle12}. Indeed similar conclusions have been found in TDEs that have been very well monitored before their peak luminosity \citep[e.g.][]{bade96,cenko12,arcavi14}. However, \src\ has shown a steady flux increase in the past decade. 

Moreover, there are three typical types of TDE X-ray spectra: 1) strong thermal emission \citep[e.g. ASASSN-14li,][]{miller15}; 2) very soft power-law continuum \citep[e.g. $\Gamma>2$ in XMMSL2~J144605.0$+$685735,][]{saxton19}; 3) hard power-law continuum \citep[e.g. $\Gamma\approx1.6$ in Swift~J164449.3$+$573451,][]{burrows11}. The spectra of \src\ do not agree with the former two types but only the third type of TDEs, which is only found in radio-loud galaxies with significant jet emission in X-rays. This is however not the case for \src\ where there is no obvious evidence for radio emission from \src\ in the VLA FIRST Survey \citep{wadadekar04}. Therefore, the increase of the mass accretion rate in \src\ is unlikely due to TDEs, or \src\ is at least not consistent with typical TDEs we have observed.

\subsection{Comparison to BH Transient State Changes}

\begin{figure}
    \centering
    \includegraphics[width=7cm]{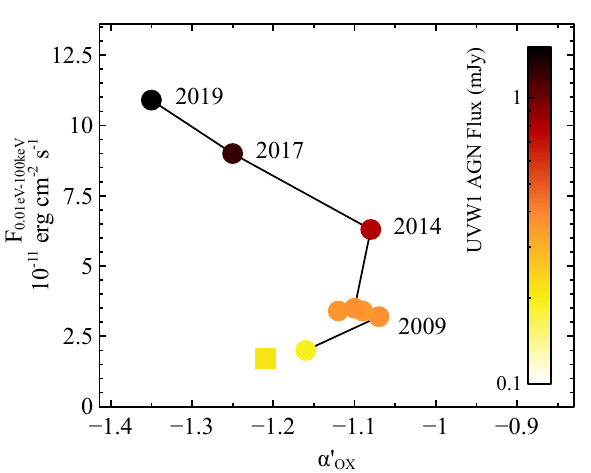}
    \caption{A broad band flux vs. X-ray--UV ratio diagram for \src. The circles and the square represent \swift\ and \xmm\ observations respectively. The color bar of the plot symbol indicates the observed UVW1 flux during the corresponding observation. }
    \label{pic_q}
\end{figure}

In this section, we discuss possible connection between \src\ and BH transients in outburst. BH transients are binary systems where the central accretors are stellar-mass BHs. The existence of two different flux states in these sources have been realised for decades \citep[e.g.][]{oda71}. Their X-ray spectra can change from a soft spectrum characterised by a strong thermal component to a hard spectrum characterised by a power-law component. Such a transition is commonly seen during an outburst of a transient and shows a `Q'-shaped pattern in the X-ray hardness-intensity diagram \citep[e.g. see a \textit{MAXI} HID of GX~339$-$4 in][]{jiang19}.

Similarly, we show the HID for \src\ in Fig.\,\ref{pic_q}. The circles in the figure represent \swift\ observations, and the square represents the \xmm\ observation in 2009. The `intensity' is the total absorption-corrected AGN flux in the 0.01\,eV--100\,keV band given by the best-fit SED models. The thermal emission from the disc in AGN is in the optical and UV bands. Therefore, we show X-ray--UV ratio, which we calculate in Section\,\ref{variability}, in the diagram as `hardness' instead of X-ray hardness as for BH transients. 

As shown in Fig.\,\ref{pic_q}, \src\ starts from the bottom right corner of the diagram in 2009 when the UV, optical and X-ray emission is all at a low flux level. Then the total luminosity of \src\ starts increasing in 2014 while the X-ray--UV ratio still remains consistent. When the source reaches the highest X-ray flux state in 2017, a lower X-ray--UV ratio is found. The latest observation in 2019 shows that the UV and optical flux is still increasing. The current state of \src\ is near the top left corner of the HID. Such a pattern is very similar to the `Q'-shaped HID of many BH transients in outburst. 

\blue{The latest \swift\ and \nustar\ observation of \src\ (obs 12) suggests a hard X-ray continuum of $\Gamma=1.7$ at a high accretion rate. The combination of a hard X-ray power-law emission and strong disc thermal emission makes obs 12 of \src\ a similar case to the intermediate state seen in some outbursts of BH transients. During these states, a stellar-mass BH transient often shows a modest mass accretion rate of $\lambda_{\rm Edd}\approx1\%$, and the disc thermal component and the non-thermal power-law component of $\Gamma<2$ make 
similar contribution to the X-ray emission \citep[e.g. GRS~1716$-$249,][]{jiang20}.}

\blue{Besides, the outbursts of BH transients can last for various timescales, e.g. weeks \citep[e.g. MAXI~1659$+$152,][]{negoro10} or years \citep[e.g. Swift~J1753.5$-$0127,][]{soleri13}. The corresponding lengths for a flaring AGN scaled by BH mass are much longer than observable timescales, e.g. $\approx10^{6}$ years. Therefore, sources like \src\ can show an outburst with a significant increase of accretion rate similar to a BH transient on very short timescales are very intriguing.}

Last but not least, a persistent radio jet is often seen during the hard state observations of BH transients \citep[e.g.][]{fender05} while a quenching of the radio emission is observed during the transition to the soft state \citep[e.g.][]{fender99}. On contrary, \src\ was not observed by the VLA FIRST survey \citep{wadadekar04}, which was taken before the transition shown in Fig.\,\ref{pic_q} started. If \src\ indeed shows a BH transient-like outburst, our hypothesis would predict simultaneous variability in the radio emission. A radio monitoring program for \src\ in future will be able to answer the question.

\subsection{Comparison to Typical Changing-Look AGN}

\blue{Changing-look AGN are rare cases of AGN, where the optical continuum flux increases or decreases and the broad emission lines appear or disappear within short time-scales. In previous studies, \citet{noda18} suggested that some `changing-look' AGN may have strong radiation or magnetic pressure in the disc, which may shorten the state transitions in AGN that are similar to BH transients. \citet{ruan19} showed that the combined $\alpha_{\rm OX}$--$\lambda_{\rm Edd}$ evolution of two `changing-look' AGN show a similar shape as \src\ does in Fig.\,\ref{pic_q} with a similar increase of $\lambda_{\rm Edd}$.} 

\blue{All of these properties make \src\ a similar case to changing-look AGN. However, it is also important to note that \src\ does not seem to change `look', e.g. Sy2 to Sy1, along with the boost of accretion rate: the \xmm\ observation taken in 2009 when the source was in a low-$\lambda_{\rm Edd}$ state was able to rule out a Compton-thick scenario that is commonly seen in Sy2 AGN \citep{risaliti99}; during the high-$\lambda_{\rm Edd}$ state, e.g. obs 12 in 2019, the X-ray spectra of \src\ still remained unobscured. Besides, the Keck observation in the optical band presented in \citet{runco16} showed that \src\ already had a typical Sy1 spectrum in 2009 during the low-$\lambda_{\rm Edd}$ state\footnote{Our follow-up optical observations taken by Lijiang Observatory during the high-$\lambda_{\rm Edd}$ state will be presented in the second paper of this series.}.}

\subsection{The Puzzling Soft Excess Emission}

We find strong evidence for soft excess below 1\,keV only in obs 12 when \src\ is in the high flux state ($\lambda_{\rm Edd}\approx3\%$). By modelling the soft excess emission with a phenomenological blackbody model (\texttt{bbody}), we obtain $kT=0.1$\,keV, which is similar to the value of a typical Sy1 \citep[e.g.][]{gierlinski99}.

\citet{noda18} suggests that a rising soft excess should be seen when an AGN goes through a `changing-look' phase with a modest mass accretion rate, e.g. $\lambda_{\rm Edd}\approx1\%$. However, we are only able to obtain an upper limit of the soft excess emission in obs 11 taken in 2017 when the X-ray flux is the highest and the continuum is the softest. This is contrary to the example of Mrk~1018 given by \citet{noda18}.

There are two most popular explanations for the origin of the soft excess emission in Sy galaxies: warm corona and relativistic disc reflection.

\begin{table*}
    \centering
    \begin{tabular}{cccccccc}
    \hline\hline
     Warm Corona Model & Parameter & Value & Reflection  Model & Parameter &  Value \\
    \hline
    \texttt{nthcomp} & $kT_{\rm e}$ (keV) & $0.17^{+0.27}_{-0.06}$ & \texttt{relconv} & q & $3^{+6}_{-2}$ \\
                     & $\Gamma_{\rm warm}$ &$2.5^{+0.2}_{-1.3}$ & & $i$ (deg) & $67^{+12}_{-42}$ \\
                     & $kT_{\rm in}$ (eV) & $3.6^{+0.2}_{-0.3}$ & & $R_{\rm in}$ ($r_{\rm g}$) & <80\\
                     & norm & $(1.4\pm1.0)\times10^{-4}$ & \texttt{reflionx} &
                     $\log(\xi$/erg\,cm\,s$^{-1}$) & $1.7^{+0.6}_{-0.2}$\\
    \texttt{simpl} & $f_{\rm scatt}$ &  $(1.9^{+1.0}_{-0.4})\times10^{-2}$ & & $kT_{\rm in}$ (eV) & $5.3\pm0.3$\\
                   & $\Gamma_{\rm hot}$ & $1.71\pm0.03$ & & $\Gamma$ & $1.72^{+0.02}_{-0.03}$ \\
    \texttt{diskbb} & $kT_{\rm in}$ & linked & & norm & $(7^{+3}_{-2})\times10^{-3}$ \\
                    & norm & $(1.2^{+0.3}_{-0.2})\times10^{10}$ & & $\log(n_{\rm e}/{\rm cm^{-3}})$ & <20 \\
      &  &                 & \texttt{simpl} & $\Gamma$ & linked\\
     &  &  & & $f_{\rm scatt}$ & $(1.2\pm0.3)\times10^{-2}$\\
                   &  &  &  \texttt{diskbb} & $kT_{\rm in}$ & linked\\
                   & & & & norm & $(4.6^{+1.2}_{-1.2})\times10^{9}$ \\
    \hline
    & $\chi^{2}/\nu$ & 492.14/404 & & $\chi^{2}/\nu$ & 486.04/401\\
    & $F_{\rm AGN, 0.01eV-100keV}$ &  12.0 & & $F_{\rm AGN, 0.01eV-100keV}$ &  10.9 \\ 
    \hline\hline
    \end{tabular}
    \caption{\blue{Best-fit warm corona and relativistic disc reflection model parameters for obs 12. The values of AGN flux are in units of $10^{-11}$ \ergps.}}
    \label{tab_warm_ref}
\end{table*}

\subsubsection{Warm corona}

In the warm corona scenario, the Comptonisation spectrum from an optically thick corona ($\tau=10-20$) with a relatively low temperature of $kT_{\rm e}<1$\,keV is used to explain the soft excess emission \citep[e.g.][]{petrucci18}. This additional corona has a much lower temperature than the `hot corona', and thus is called `warm corona'. \blue{In order to test for this model, we calculate the warm coronal emission by using the Comptonisation model \texttt{nthcomp}. A disc blackbody-shaped seed photon spectrum is considered in consistency with the thermal disc component \texttt{diskbb}. The temperature parameter for the seed photon spectrum in \texttt{nthcomp} is linked to the corresponding parameter of the \texttt{diskbb} model. The total model is \texttt{tbnew * zdust * (nthcomp + simpl*diskbb)}. The best-fit parameters are shown in Table\,\ref{tab_warm_ref}, and the best-fit model is shown in the top left panel of Fig.\,\ref{pic_soft}. Corresponding data/model ratio plots are shown in the bottom left panel of Fig.\,\ref{pic_soft}.}

\blue{According to our warm corona-based SED model, the extreme UV emission (0.02\,eV--0.2\,keV) of \src\ is dominated by the warm coronal emission. Consequently, a higher bolometric luminosity is predicted by this model ($F_{\rm AGN}=1.2\times10^{-10}$\,\ergps\ in the 0.01\,eV--100\,keV band corresponding to an Eddington ratio of $\lambda_{\rm Edd}\approx3.6\%$).} 

\blue{Similar conclusions are achieved in other warm corona-based analyses of narrow-line Sy1s \citep[e.g.][]{jin17} and `changing-look' AGN \citep[e.g.][]{noda18}, where they also find the emission at longer wavelength dominated by warm coronal emission. For instance, \citet{noda18} modelled the variable soft excess in the `changing-look' AGN Mrk~1018 with warm corona models, and concluded that the phase transition between Sy1 and Sy1.9 in Mrk~1018 is controlled by the variable warm Comptonisation region. It is interesting to note that, in comparison, \src\ only shows evidence for weak soft excess in obs 12 but not in obs 11 when \src\ is in the highest and the softest X-ray state and a high optical and UV flux state. We do not expect the extreme UV emission to switch completely between a thermal-dominant state to a non-thermal/warm corona-dominant state without much flux and spectral change as suggested by the warm corona model.}

\subsubsection{Disc reflection}

\blue{In the disc reflection scenario, the soft excess is explained by reflection from the innermost region of the disc \citep[e.g.][]{crummy06,walton13, jiang19d} within only 10-20 $R_{\rm g}$ from the BH \citep[e.g.][]{morgan08, reis13, chartas17, fabian20}. The decreasing inner radius inferred by our SED models in Section\,\ref{sed_disc} also suggests a thin disc that is forming in the region very close to the ISCO. The soft excess may arise as part of the reflection spectrum from the inner disc.}

\begin{figure*}
    \centering
    \includegraphics{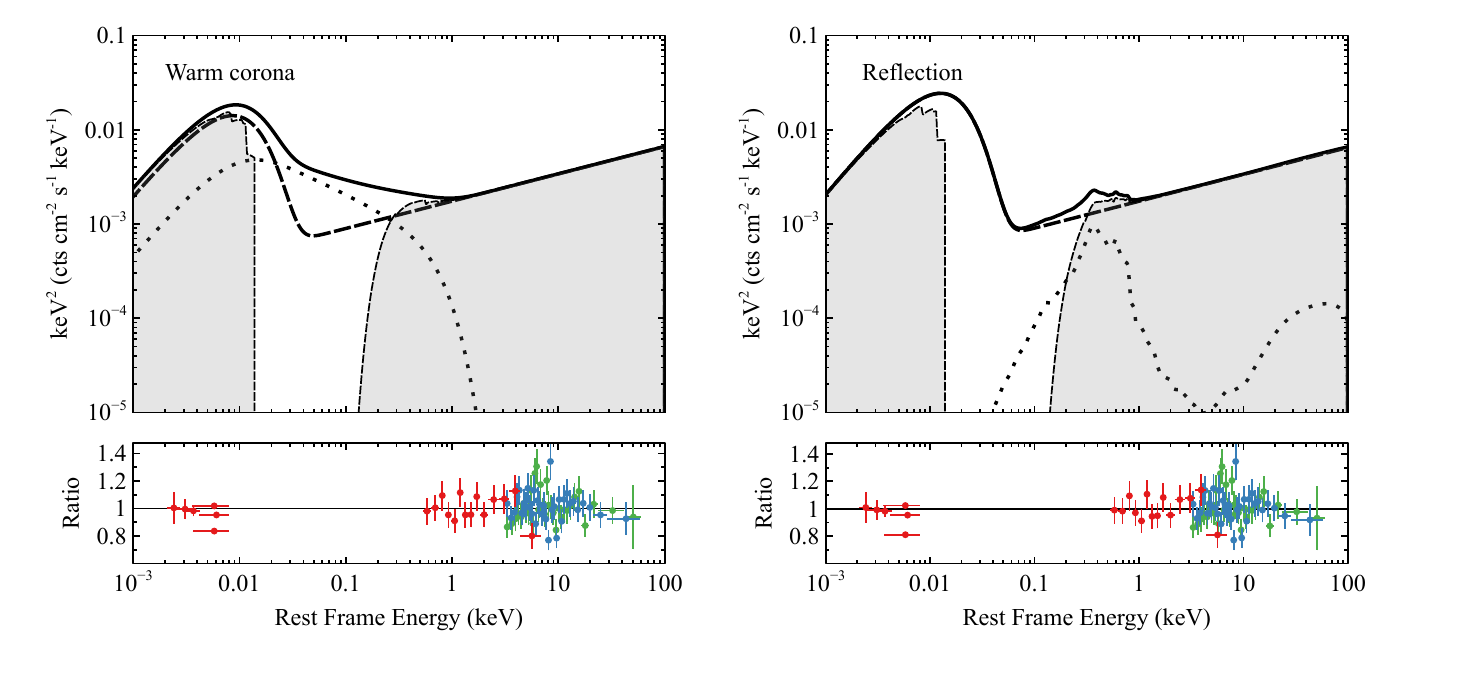}
    \caption{\blue{Top panels: best-fit warm corona-based (left) and reflection-based (right) SED models for obs 12. Grey shaded regions: total SED models; black solid lines: unabsorbed total models; the dotted line in the left panel: the warm corona emission (\texttt{nthcomp}); the dotted line in the right panel: the best-fit relativistic disc reflection component; dashed lines: the disc thermal emission and the hot coronal emission (\texttt{simplcut*diskbb}). The best-fit warm corona model suggests that the extreme UV emission from the AGN of \src\ is dominated by warm coronal emission. Bottom panels: data/model ratio plots using the corresponding SED model in the top panel.}}
    \label{pic_soft}
\end{figure*}

\blue{In order to test for reflection models, we replace the \texttt{bbody} model with an extended version of the ionised plasma model \texttt{reflionx} \citep{ross93}. The illuminating spectrum of \texttt{reflionx} is calculated using \texttt{nthcomp}. A variable disc seed photon temperature is considered and linked to the inner temperature of the disc $kT_{\rm in}$ \citep{jiang20b}. The convolution model \texttt{relconv} is used to account for relativistic effects \citep{dauser13}. The best-fit parameters are shown in the last column of Table\,\ref{tab_warm_ref}, and the best-fit model is shown in the top right panel of Fig.\,\ref{pic_soft}. Corresponding data/model ratio plots are shown in the lower panel.}

\blue{The relativistic reflection model provides an equally good fit as the warm corona model with $\Delta\chi^{2}=6$ and 3 more free parameters. The inclination angle of the disc is consistent with either a face-on ($i\approx25^{\circ}$) or a high-inclination ($i\approx79^{\circ}$) scenario within a 90\% confidence uncertainty range. Due to the limited signal-to-noise, we are only able to obtain an upper limit of the inner radius of the disc ($R_{\rm in}<80$\,$r_{\rm g}$), which is statistically consistent with measurements in Section\,\ref{sed_disc}.}

\blue{In comparison with the warm corona model, the reflection model indicates stronger disc thermal emission in the optical and UV band (see the top two panels in Fig.\,\ref{pic_soft}). The properties of the thermal disc component and the up-scattering fraction in the hot coronal region are consistent with the results obtained by modelling the soft excess with a blackbody model as in Section\,\ref{sed}. A similar Eddington ratio of $\lambda_{\rm Edd}\approx3.2\%$ is also obtained.}

\blue{The apparent absence of broad Fe~K emission line in the data does not rule out of the disc reflection scenario especially when the soft excess emission is shown \citep[e.g.][]{gallo06,garcia19,jiang20b}. They might be related to, for example, 1) a low reflection fraction ($f_{\rm refl}$); 2) strong relativistic effects; 3) certain properties of the disc, e.g. ionisation and density; 4) limited signal-to-noise of the data in the iron band.}

\blue{In the case of \src, it is important to note that our best-fit reflection model suggests a low-$f_{\rm refl}$ scenario during obs 12 of \src\ despite the clear evidence of soft excess emission. For instance, the reflection component only takes 2\% of the total flux in the iron band (4--8\,keV). A similar $f_{\rm refl}$ is shown in the hard X-ray band (20--30\,keV), where the Compton hump is shown. Therefore, the lack of clear evidence for broad Fe K emission line is possibly caused by the low-$f_{\rm refl}$ nature of the source during obs 12 and the limited signal-to-noise of the data. Observations by future missions in soft X-rays, e.g. \textit{Athena}, in combination with high-sensitivity hard X-ray mission, e.g. \textit{HEX-P}, will provide a unique opportunity to constrain the broad Fe K emission line and the Compton hump in a low-$f_{\rm refl}$ scenario as in \src.}

\section{Conclusions}

In this work, we present multi-epoch X-ray spectral analysis and SED modelling of a Sy1 called \src. \src\ shows a simultaneous steady flux increase in the optical and UV bands since 2009. For instance, the UVW1 flux of the AGN in \src\ has increased by more than one order of magnitude during the latest \swift\ observation on 26th December, 2019 compared to the observation in 2009. It is interesting that the optical--UV continuum becomes flatter at a higher flux state, which can be explained by an increase of the temperature and the accretion rate in the disc. 

The X-ray luminosity of \src\ shows a steady increase by one order of magnitude from 2007 to 2017 as well. The broad band HID of \src\ in the last decade is very similar to the X-ray HID of stellar-mass BH transients in outburst. Such a rapid boost of mass accretion rate makes \src\ a very interesting source. \blue{By modelling multi-wavelength SEDs, the luminosity of \src\ increases from $0.6\%$ of the Eddington limit in 2009 to 3.2\% in 2019.} Along with the rapid increase of mass accretion rate, a possible decreasing inner radius of the disc is suggested by our SED models. We only find obvious evidence of soft excess in the latest observation in 2019, during which \src\ has an Eddington ratio of $>3\%$ and the UV and optical flux is shown to be the highest. 

\section*{DATA AVAILABILITY}

The data underlying this article are available in the High Energy Astrophysics Science Archive Research Center (HEASARC), at https://heasarc.gsfc.nasa.gov.

\section*{Acknowledgements}

This paper was written during the outbreak of COVID-19 in China in 2020. We would like to acknowledge the doctors and nurses who have been working day and night to ensure the safety of Chinese people during this period. JJ acknowledges support from the Tsinghua Astrophysics Outstanding Fellowship and the Tsinghua Shuimu Scholar Program. LCH acknowledges support from National Science Foundation of China (11721303, 11991052) and National Key R\&D Program of China (2016YFA0400702). DJKB acknowledges support from Royal Society. ACF acknowledges support from ERC Advanced Grant (340442). CSR thanks the UK Science and Technology Facilities Council for support under the New Applicant grant ST/R000867/1, and the European Research Council for support under the European Union’s Horizon 2020 research and innovation programme (834203). MLP is supported by European Space Agency Research Fellowships. DJW acknowledges support from an STFC Ernest Rutherford Fellowship. This work made use of data from the \nustar\ mission, a project led by the California Institute of Technology, managed by the Jet Propulsion Laboratory, and funded by NASA. This research has made use of the \nustar\ Data Analysis Software (NuSTARDAS) jointly developed by the ASI Science Data Center and the California Institute of Technology. This work made use of data supplied by the UK Swift Science Data Centre at the University of Leicester. This project was also based on observations obtained with \xmm, an ESA science mission with instruments and contributions directly funded by ESA Member States and NASA. This project has made use of the Science Analysis Software (SAS), an extensive suite to process the data collected by the \xmm\ observatory.

%%%%%%%%%%%%%%%%%%%%%%%%%%%%%%%%%%%%%%%%%%%%%%%%%%



\bibliographystyle{mnras}
\bibliography{kug1141.bib} % if your bibtex file is called example.bib


%%%%%%%%%%%%%%%%%%%%%%%%%%%%%%%%%%%%%%%%%%%%%%%%%%

%%%%%%%%%%%%%%%%% APPENDICES %%%%%%%%%%%%%%%%%%%%%

\appendix

\section{Error estimation on the UVOT photometric measurements}
\label{sec:uvoterror}

\blue{Here we briefly introduce our methods on the error estimation for the UVOT photometric measurements. Given that the image decomposition procedures are only implemented on the optical images, we take two different strategies for UV and optical data respectively.}

\blue{For the photometric uncertainties in three UV filters (UVW1, UVM2, and UVW2), we simply adopt the measurements returned by the \textsc{UVOTSOURCE} tool, including both systematic and statistical errors. The systematic errors mainly refer to the uncertainties on the zero-point and the conversion from counts to flux (depending on the assumed spectral shape), and are added in quadrature to the statistical errors in the calculation of the total errors \citep{add08,add10b}.}

\blue{The estimation of the uncertainties for AGN fluxes at longer wavelengths (V, B, and U filters) are addressed in a different way, as the errors in the image decomposition procedure needs to be taken into account. One of the highlights in our work is that a similar image decomposition procedure was carried out on the same object in multiple observations. Therefore, it is doable to obtain an error of the host galaxy flux from (some of) these measurements and assess the error of the AGN flux through the `error propagation formula'.}

\blue{A straightforward relation between the observed flux is that
\begin{equation}
\label{eq:fluxall}
F_{\rm AGN} = F_{\rm all}-F_{\rm gal}-F_{\rm back},
\end{equation}
in which $F_{\rm all}$ is the flux for the whole galaxy, including the AGN emission $F_{\rm AGN}$, the host starlight $F_{\rm gal}$ and the background $F_{\rm back}$. With the error propagation formula we can get
\begin{equation}
\label{eq:error}
\sigma(F_{\rm AGN}) = \sqrt{\sigma(F_{\rm all})^2+\sigma(F_{\rm gal})^2},
\end{equation}
note that $\sigma(F_{\rm back})$ can be neglected. Therefore, $\sigma(F_{\rm AGN})$ can be obtained by calculating $\sigma(F_{\rm all})$ and $\sigma(F_{\rm gal})$.}

\blue{$\sigma(F_{\rm gal})$ is estimated from the multiple image fitting results. We can get the averaged flux density and its standard deviation at the (observed) frequency $\nu$, by}

\blue{
\begin{equation}
\overline{F_{\nu, \rm gal}} = \frac{\sum\limits_{i=1}^{N}F_{\nu, \rm gal}(i)}{N}
\end{equation}
\begin{equation}
\sigma(\overline{F_{\nu, \rm gal}}) = \sqrt{\frac{\sum\limits_{i=1}^{N}\left(F_{\nu, \rm gal}(i)-\overline{F_{\nu, \rm gal}}\right)^2}{N\times (N-1)}} 
\end{equation}
}
\blue{Table \ref{tab:host} lists the photometry for the host galaxy starlight for the eight UVOT observations in three optical bands, as well as the averaged flux densities and 1$\sigma$ errors.}

\blue{$\sigma(F_{\rm all})$ is estimated by using the \textsc{UVOTSOURCE} tool. Compared to the aperture utilized in the UV bands (5 arcsec), a larger aperture of 15 arcsec is employed. The results are listed in Table \ref{tab:error_all}. With Equation \ref{eq:error}, the errors of the AGN emission in three optical bands can be obtained. The photometric uncertainties utilized in the spectral fitting are summarized in Table \ref{tab:final_agn}, containing those in all the six UVOT filters.}

\blue{One thing should be pointed out that above method for the error estimation of the optical data is based on the premise that the model employed in the image fitting, i.e. `PSF + exponential-disc + background' is the {\it true} model for the description of the light distribution of KUG 1141. Therefore, the estimated errors does not contain those arising from the deviations of the model from the reality, i.e. an extra component may need to be included, such as a bulge/bar. Due to the low spatial resolution of the UVOT images, to thoroughly address this issue is beyond the scope of this paper.}

\begin{table}
\centering
\caption{\blue{The estimated flux density for the host galaxy in three optical bands. The flux densities are in units of {\it mJy}. For U band, we do not take the results for the second UVOT observation (Obs ID. 00037565002) due to the poor data quality (the parameters in the exponential disc cannot be well constrained).}}
\label{tab:host}
\small
\renewcommand\arraystretch{1.0}
\begin{tabular}{ccccccccccc} % four columns, alignment for each
\hline
\hline
Obs No. & $F_{\nu, \rm gal}$, V & $F_{\nu, \rm gal}$, B & $F_{\nu, \rm gal}$, U  \\
\hline
4 & 2.95 & 1.44 & 0.35 \\
6 & 2.50 & 1.67 & - \\
7 & 2.42 & 1.57 & 0.37 \\
8 & 2.10 & 1.10 & 0.25 \\
9 & 2.41 & 1.25 & 0.38 \\
10 & 2.15 & 1.17 & 0.28 \\
11 & 2.68 & 1.18 & 0.38 \\
12 & 2.72 & 1.25 & 0.36  \\
\hline
$\overline{F_{\nu, \rm gal}}$ & 2.49 & 1.33 & 0.34 \\
$\sigma(\overline{F_{\nu, \rm gal}})$ & 0.10 & 0.07 & 0.02 \\
\hline
\end{tabular}
\end{table}

\begin{table}
\centering
\caption{\blue{The standard deviation of the AGN+host galaxy emission flux density, $\sigma(F_{\nu, \rm all})$, in the three optical bands for different UVOT observations. The flux densities are in units of {\it mJy}.}}
\label{tab:error_all}
\small
\renewcommand\arraystretch{1.0}
\begin{tabular}{ccccccccccc} % four columns, alignment for each
\hline
Obs No. & $\sigma(F_{\nu, \rm all})$, V & $\sigma(F_{\nu, \rm all})$, B & $\sigma(F_{\nu, \rm all})$, U \\
\hline
4 & 0.11 & 0.05 & 0.02 \\
6 & 0.13 & 0.06 & 0.03 \\
7 & 0.16 & 0.07 & 0.04 \\
8 & 0.12 & 0.06 & 0.03 \\
9 & 0.11 & 0.05 & 0.03 \\
10 & 0.15 & 0.07 & 0.03 \\
11 & 0.14 & 0.07 & 0.05 \\
12 & 0.15 & 0.08 & 0.06 \\
\hline
\hline
\end{tabular}
\end{table}

\begin{table*}
\caption{\blue{The AGN flux density utilized in the spectral fitting, $F_{\rm AGN}$, in all six UVOT filters. The flux densities are in units of {\it mJy}.}} \label{tab_uvot}
\begin{tabular}{ccccccc}
\hline\hline
Obs No. & $F_{\nu, \rm AGN}$, V & $F_{\nu, \rm AGN}$, B & $F_{\nu, \rm AGN}$, U & $F_{\nu, \rm AGN}$, UVW1 & $F_{\nu, \rm AGN}$, UVM2 & $F_{\nu, \rm AGN}$, UVW2 \\
\hline
4       & 0.78 & 0.47 & 0.25 & 0.19 & 0.13 & 0.12 \\
6       & 0.74 & 0.45 & 0.49 & 0.37 & 0.27 & 0.24 \\
7       & 0.88 & 0.46 & 0.42 & 0.35 & 0.27 & 0.24 \\
8       & 0.64 & 0.51 & 0.48 & 0.36 & 0.26 & 0.24 \\
9       & 0.88 & 0.47 & 0.48 & 0.37 & 0.26 & 0.23 \\
10      & 1.26 & 1.14 & 1.07 & 0.79 & 0.62 & 0.56 \\
11      & 1.35 & 1.21 & 1.45 & 1.24 & 1.07 & 1.01 \\
12      & 1.59 & 1.53 & 1.64 & 1.59 & 1.21 & 1.23 \\
\hline\hline
\end{tabular}
\end{table*}

\begin{table*}
\centering
\caption{\blue{The standard deviation of the AGN flux density utilized in the spectral fitting, $\sigma(F_{\rm AGN})$, in all six UVOT filters. The flux densities are in units of {\it mJy}.}}
\label{tab:final_agn}
\small
\renewcommand\arraystretch{1.0}
\begin{tabular}{ccccccccccc} % four columns, alignment for each
\hline\hline
Obs No. & $\sigma(F_{\nu, \rm AGN})$, V & $\sigma(F_{\nu, \rm AGN})$, B & $\sigma(F_{\nu, \rm AGN})$, U & $\sigma(F_{\nu, \rm AGN})$, UVW1 & $\sigma(F_{\nu, \rm AGN})$, UVM2 & $\sigma(F_{\nu, \rm AGN})$, UVW2 \\
\hline
4 & 0.15 & 0.09 & 0.03 & 0.01 & 0.01 & 0.01 \\
6 & 0.17 & 0.09 & 0.04 & 0.02 & 0.01 & 0.01 \\
7 & 0.19 & 0.10 & 0.04 & 0.01 & 0.02 & 0.01 \\
8 & 0.16 & 0.09 & 0.03 & 0.02 & 0.01 & 0.01 \\
9 & 0.15 & 0.09 & 0.03 & 0.02 & 0.01 & 0.01 \\
10 & 0.18 & 0.10 & 0.04 & 0.04 & 0.02 & 0.02 \\
11 & 0.17 & 0.10 & 0.05 & 0.05 & 0.03 & 0.03 \\
12 & 0.18 & 0.11 & 0.06 & 0.06 & 0.04 & 0.04 \\
\hline
\hline
\end{tabular}
\end{table*}

\section{Additional Information of X-ray and SED Analysis} \label{extra}

\begin{figure}
    \centering
    \includegraphics{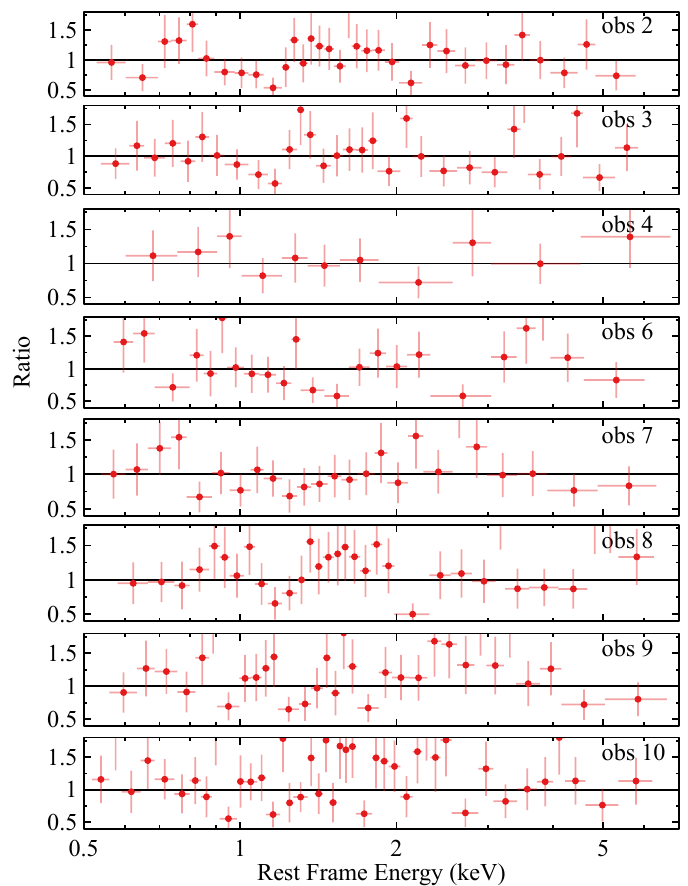}
    \caption{Data/model ratio plots for \swift\ XRT spectra of \src\ using an absorbed power-law model. The ratio plots for obs 1, 5, 11 and 12 can be found in Fig.\,\ref{pic_obs5},\ref{pic_obs12} and \ref{pic_obs1}}
    \label{pic_xrt_ratio}
\end{figure}

Detailed X-ray flux extracted from all the epochs is shown in Table\,\ref{tab_flux} and visulised in Fig.\,\ref{pic_lc}. Data/model ratio plots for the \swift\ observations of \src\ using an absorbed power law are shown in Fig.\,\ref{pic_xrt_ratio}. The best-fit SED parameters for obs 4--12 are listed in Fig.\,\ref{tab_sed} and visilised in Fig.\,\ref{pic_sed_fit}.

\begin{table}
    \centering
    \begin{tabular}{ccccccccccccccccc}
    \hline\hline
    Obs No. & $\log(F_{\rm 0.5-7keV})$ & $\log(F_{\rm 0.5-2keV})$ & $\log(F_{\rm 2-7keV})$  \\
    \hline
      1 & $-11.79\pm0.06$ & $-12.22\pm0.06$ & $-11.99\pm0.09$ \\
      2 & $-11.79\pm0.05$ & $-12.23\pm0.05$ & $-11.99\pm0.07$ \\
      3 & $-11.71\pm0.05$ & $-12.19\pm0.05$ & $-11.89^{+0.07}_{-0.08}$ \\
      4 & $-11.49\pm0.08$ & $-11.89\pm0.08$ & $-11.72^{+0.13}_{-0.16}$ \\
      5 & $-11.616^{+0.015}_{-0.006}$ & $-11.997\pm0.012$ & $-11.85\pm0.02$ \\
      6 & $-11.25\pm0.06$ & $-11.62\pm0.05$ & $-11.49\pm0.09$ \\
      7 & $-11.22\pm0.06$ & $-11.62\pm0.05$ & $-11.45\pm0.09$ \\
      8 & $-11.35^{+0.05}_{-0.06}$ & $-11.79\pm0.05$ & $-11.54\pm0.09$ \\
      9 & $-11.35\pm0.05$ & $-11.76\pm0.05$ & $-11.57\pm0.07$ \\
      10 & $-11.06^{+0.04}_{-0.05}$ & $-11.46\pm0.04$ & $-11.28\pm0.08$ \\
      11 & $-10.92\pm0.06$ & $-11.22\pm0.06$ & $-11.23^{+0.11}_{-0.12}$ \\
      12 & $-11.08\pm0.02$ & $-11.43\pm0.02$ & $-11.33\pm0.04$ \\
      
    \hline\hline
    \end{tabular}
    \caption{Observed X-ray flux values in units of erg\,cm$^{-2}$\,s$^{-1}$.}
    \label{tab_flux}
\end{table}

\begin{table*}
    \centering
    \begin{tabular}{ccccccccc}
    \hline\hline
       Obs & $kT_{\rm in}$ &  \texttt{diskbb} norm & $\Gamma$ & $f_{\rm scatt}$ & $F_{\rm tol}^{1}$ &  $\chi^{2}/\nu$\\
       No. & eV & $10^{10}$  & - & $10^{-2}$ & $10^{-11}$ \ergps  \\
    \hline
    4 & $2.0\pm0.2$ & $2.6^{+0.8}_{-0.6}$ & $1.7\pm0.2$ & $2.7^{+0.5}_{-0.4}$ & 2.0 &  19.13/13 \\
    5 & $2.1^{+0.2}_{-0.3}$ & $1.9^{+0.4}_{-0.3}$ & $1.70\pm0.04$ & $2.1^{+0.3}_{-0.4}$ & 1.7&  157.87/164\\
    6 & $2.8\pm0.3$ & $0.8^{+0.3}_{-0.2}$ & $1.70\pm0.04$ & $4.9^{+0.6}_{-0.5}$ & 3.2 & 44.68/25\\
    7 & $2.5^{+0.3}_{-0.2}$ & $1.2^{+0.6}_{-0.4}$ & $1.67\pm0.18$ & $4.0^{+0.5}_{-0.4}$ & 3.4 & 27.59/27\\
    8 & $2.9^{+0.3}_{-0.2}$ & $0.7\pm0.2$ & $1.52\pm0.18$ & $2.2\pm0.2$ & 3.4 & 48.23/34\\
    9 & $2.8^{+0.3}_{-0.2}$ & $0.7^{+0.3}_{-0.2}$ & $1.62\pm0.06$ & $2.7^{+0.3}_{-0.2}$ & 3.5 & 49.22/34\\
    10 & $2.7\pm0.1$ & $0.8^{+0.4}_{-0.3}$ & $1.65\pm0.11$ & $2.8\pm0.2$ & 6.3 & 60.88/44\\
    11 & $4.8^{+0.5}_{-0.4}$ & $0.5^{+0.4}_{-0.3}$ & $2.0\pm0.2$ & $9.3^{+0.5}_{-0.4}$ & 9.0 & 43.55/19\\
    12 & $5.3^{+0.4}_{-0.3}$ & $0.5\pm0.2$ & $1.71\pm0.03$ & $1.5\pm0.2$ & 10.9 & 489.81/405\\
    \hline\hline
    \end{tabular}
    \caption{\blue{The best-fit SED model parameters for obs 4-12. $^{1}$ The flux of the AGN emission is calculated using the best-fit model in the 0.01\,eV--100\,keV band.}}
    \label{tab_sed}
\end{table*}

\section{Short-Term Variability on Kilo-second Timescales} \label{short}

\begin{figure}
    \centering
    \includegraphics[width=\columnwidth]{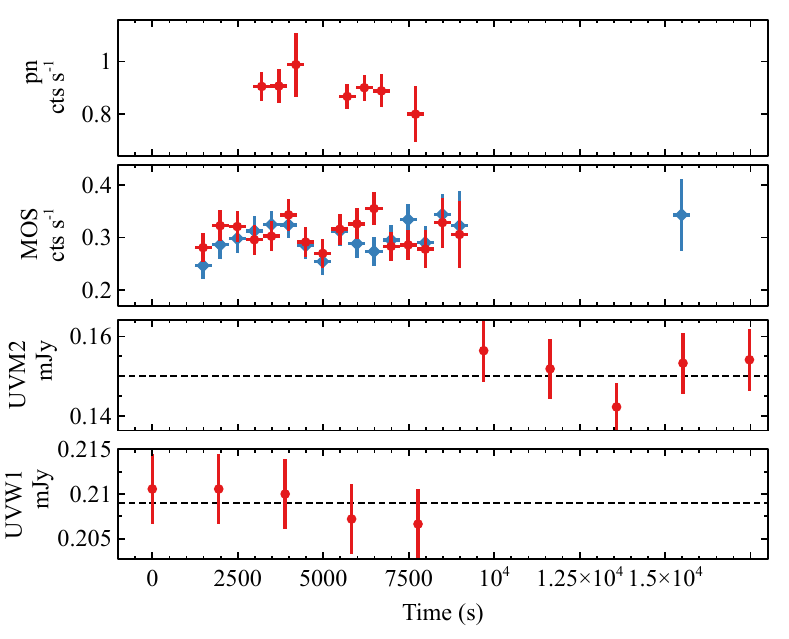}
    \caption{X-ray (0.5--10\,keV, top two panels) and UV (UVW1, UVM2, bottom two panels) lightcurves extracted from the archival \xmm\ observation (obs 5) of \src. The gaps in the X-ray lightcurves are due to high flaring particle background. The X-ray lightcurves are in 500s bin. The red crosses in the second panel show the MOS1 light curve, and the blue crosses show the MOS2 light curve. The UV light curves show the flux given by each OM exposure. The dashed lines in the bottom panels show the mean observed flux during this observation.}
    \label{pic_short_variability}
\end{figure}

\begin{figure}
    \centering
    \includegraphics[width=\columnwidth]{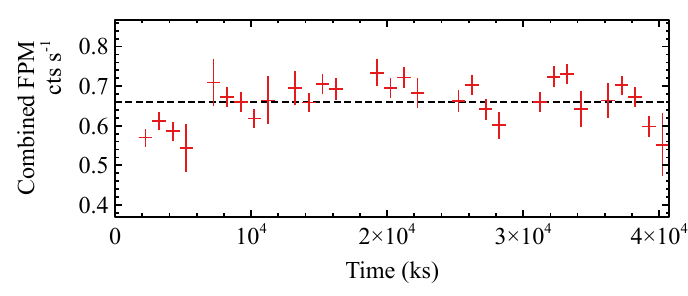}
    \caption{A 0.5--60\,keV lightcurve extracted from the \nustar\ (obs 12) of \src. The dashed line shows the averaged combined FPM count rate.}
    \label{pic_fpm_lc}
\end{figure}

Fig.\,\ref{pic_short_variability} and Fig.\,\ref{pic_fpm_lc} show the \xmm\ and \nustar\ lightcurves of \src extracted from obs 5 and 12. The gaps in the EPIC lightcurves are due to high flaring particle background. Both the X-ray and UV emission remains at a consistent flux level within this observation. This suggests that the multi-wavelength flux increase shown in Fig.\,\ref{pic_lc} is not due to the lack of frequent monitoring of a very fast intrinsic AGN variability on timescales of kiloseconds.


%%%%%%%%%%%%%%%%%%%%%%%%%%%%%%%%%%%%%%%%%%%%%%%%%%

% Don't change these lines
\bsp	% typesetting comment
\label{lastpage}
\end{document}